\documentclass[10pt, conference, letterpaper]{IEEEtran}

\makeatletter
\def\ps@headings{%
\def\@oddhead{\mbox{}\scriptsize\rightmark \hfil \thepage}%
\def\@evenhead{\scriptsize\thepage \hfil \leftmark\mbox{}}%
\def\@oddfoot{}%
\def\@evenfoot{}}
\makeatother
\author{}
\usepackage{url}
\usepackage{enumerate}
\newcommand{\compactlist}{\setlength{\itemsep}{0pt} \setlength{\parskip}{0pt} \setlength{\leftskip}{-1em}}
\usepackage{cite}
\usepackage{setspace}
\usepackage{soul}

\usepackage{multirow}
\usepackage[linesnumbered,ruled]{algorithm2e}
\usepackage[T1]{fontenc}
\usepackage{graphics}
\usepackage{amsmath}
\usepackage{amsfonts}
\usepackage{amssymb}
\usepackage{epstopdf}
\include{psfig}
\usepackage{epsfig}
\usepackage{xspace}
\usepackage{amssymb}
\usepackage{subfigure}
\usepackage{booktabs}

\usepackage{mathtools}
\usepackage{bm} 
\usepackage{graphicx}

\usepackage{framed} 
\usepackage[framed]{ntheorem}

\newframedtheorem{frm-thm}{Theorem}
\newframedtheorem{frm-col}{Corrolary}
\newframedtheorem{frm-pol}{Policy}

\newtheorem{definition}{Definition}

\newframedtheorem{conjecture}{Conjecture}

\newenvironment{proof}{{\bf{Proof:}}}{\hfill$\blacksquare$} 

\newcommand \be {\begin{equation}}
\newcommand \bes {\begin{equation*}}

\newcommand \ee {\end{equation}}
\newcommand \ees {\end{equation*}}

\newcommand \M {\begin{bf} M \end{bf}}

\newcommand{\ie}{\textit{i.e.}\xspace}

\newcommand{\eg}{\textit{e.g.}\xspace}

\usepackage{tikz}
\newcommand*\circled[1]{\tikz[baseline=(char.base)]{
            \node[shape=circle,draw,inner sep=2pt] (char) {#1};}}

\newcommand\EatDot[1]{}

\newcommand{\w}{\omega}

\newcommand{\thr}{\it{r}}
\newcommand{\ttheta}{{\theta}}

\newcommand{\blambda}{\bm{\lambda}}

\newcommand \N {\begin{bf} N \end{bf}}

\newcommand{\bbr}{{\bf{r}}}
\newcommand{\bmu}{\pmb{\mu}}
\newenvironment{proof-sketch}{\noindent{\bf Sketch of Proof:}\hspace*{1em}}{\qed\bigskip}
\newenvironment{proof-E}{\noindent{\bf Proof:}\hspace*{1em}}{\qed\bigskip}

\makeatletter
\renewcommand{\p@subfigure}{\thefigure}
\makeatother

\begin{document}


\title{Proportional Fair RAT Aggregation in HetNets}
\author{\IEEEauthorblockN{Ehsan Aryafar}
\IEEEauthorblockA{Portland State University\\
Portland, OR}  \and
\IEEEauthorblockN{Alireza Keshavarz-Haddad}%
\IEEEauthorblockA{Shiraz University\\
Shiraz, Iran} \and
\IEEEauthorblockN{Carlee Joe-Wong}%
\IEEEauthorblockA{Carnegie Mellon University\\
Silicon Valley, CA \\}}
\maketitle

\begin{abstract}
Heterogeneity in wireless network architectures (\ie, the coexistence of 3G, LTE, 5G, WiFi, $etc.$) has become a key component of current and future generation cellular networks. Simultaneous aggregation of each client's traffic across multiple such radio access technologies (RATs) / base stations (BSs) can significantly increase the system throughput, and has become an important feature of cellular standards on multi-RAT integration. Distributed algorithms that can realize the full potential of this aggregation are thus of great importance to operators. In this paper, we study the problem of resource allocation for multi-RAT traffic aggregation in HetNets (heterogeneous networks). Our goal is to ensure that the resources at each BS are allocated so that the aggregate throughput achieved by each client across its RATs satisfies a proportional fairness (PF) criterion. In particular, we provide a simple distributed algorithm for resource allocation at each BS that extends the PF allocation algorithm for a single BS. Despite its simplicity and lack of coordination across the BSs, we show that our algorithm converges to the desired PF solution and provide (tight) bounds on its convergence speed. We also study the characteristics of the optimal solution and use its properties to prove the optimality of our algorithm's outcomes.
\end{abstract}

\section{Introduction}

%
The increasing demand for wireless data has led to denser and more heterogeneous wireless network deployments.  
This heterogeneity manifests itself in terms of network deployments across multiple radio access technologies (\eg, 3G, LTE, WiFi, 5G), cell sizes (\eg, macro, pico, femto), and frequency bands (\eg, TV bands, 1.8-2.4 GHz, mmWave), $etc$. To realize the gains associated with such heterogeneous networks (HetNets), consumer (client) devices are also being equipped with an increasing number of radio access technologies (RATs), and some are already able to simultaneously aggregate the traffic across multiple RATs to increase throughput~\cite{SAMSUNG}. 


To support such traffic 
aggregation on the network side, the 3GPP (3rd generation partnership project) has been actively developing 
multi-RAT integration solutions. 
The introduction of LWA (LTE-WiFi Aggregation) as part of the 3GPP Release 13~\cite{3GPPLWA} was a step in this direction. LWA allows using both LTE and WiFi links for a single traffic flow and is generally more efficient than transport layer aggregation protocols (\eg, MultiPath TCP), 
due to coordination at lower protocol stack layers. LWA's design primarily follows the LTE Dual Connectivity (DC) architecture (defined in 3GPP Release 12~\cite{LTEDC}), which allows a wireless device to connect to two LTE eNBs that are on different carrier frequencies, and utilize the radio resources that belong to both of them. Currently, the 3GPP is working on a solution to support below IP (layer 2) multi-RAT integration across {\it{any}} combination of RATs, including LTE, WiFi, 802.11ad/ay, and 5G New Radio (NR)~\cite{3GPPNR}. The proposed architecture would allow for dynamic  traffic splitting across RATs for each client, which can lead to a significant 
increase in the system performance (\eg, total throughput). 

However, it is difficult to design resource allocation algorithms for each BS\footnote{{\bf{We use ``BS'' generically to mean an LTE eNB, WiFi AP, etc.}}} 
that realize the performance benefits of such integrated HetNets. 
Specifically, {\bf{(i)}} backhaul links from different BSs in HetNets show diverse capacity and latency characteristics  and depend on the underlying backhauling technology. For example, cable and DSL have on average 28 and 62 ms roundtrip latencies, respectively~\cite{FCC1, FCC2}. The latency can be even higher when a network operator uses a third party ISP to communicate with its BSs (\eg, a mobile operator that uses a wired ISP to control its WiFi BSs).  
%
Such latencies make it infeasible for BSs to communicate with each other or a central controller for real-time resource allocation at each BS. As a result, any practical resource allocation algorithm for multi-RAT HetNets should be fully distributed (\ie, autonomously executed by each BS). {\bf{(ii)}} Resource allocation has many practical constraints. Conventional BS hardware allows only minor modifications to existing resource allocation algorithms through software updates, limiting the algorithm design space. New algorithms should also incur minimal signaling overhead and computational complexity.
%
Distributed algorithms based on the traditional network utility maximization framework~\cite{MUNGNUM, KELLY} do not meet these requirements, because as we will show later through simulations the resulting algorithms are radically different from how conventional BSs operate, have significant over-the-air signaling overhead, and increase the computational complexity on the client side. {\bf{(iii)}} In HetNets, each client has access to a client-specific set of RATs, and receives packets at a different PHY rate on each RAT. These rates are naturally different across clients. This multi-rate property of HetNets makes it particularly challenging to design resource allocation algorithms with performance guarantee. As a result, existing solutions in the literature are all limited to simple setups, \eg, when each client has only two RATs as in the case of LWA~\cite{sarabjotPF} or LTE DC~\cite{Prasad2017}. 

In this paper, we study the problem of resource allocation for traffic aggregation in multi-RAT HetNets. We focus on the {\it{proportional-fair (PF)}} fairness objective as it is widely used and implemented in BSs and provides a balance between fairness and throughput~\cite{propfair1, propfair2}. We first consider 
PF resource allocation in a single BS, and then use our insights from this case to design a distributed algorithm that meets our three research challenges. We next show that our algorithm converges to an optimal PF resource allocation. 
The key contributions are as follows:

\begin{itemize}
\compactlist
\item {\bf{Algorithm Design:}} We study the basics of PF resource allocation in a single BS to gain intuition for the distributed algorithm design. 
We show that PF resource allocation in a single BS can be viewed as a special type of water-filling. We generalize this observation to a new fully distributed water-filling algorithm (named AFRA) that makes a minor modification to the conventional single BS algorithm and achieves PF in HetNets.

\item {\bf{Convergence and Speed:}} 
We show that AFRA is guaranteed to converge to an equilibrium as BSs autonomously execute it [Theorem 1] and derive tight bounds on its convergence time (speed) [Theorem~2]. 
%


\item {\bf{Optimality:}} 
We first show that at optimality, the sum of the inverse water-fill levels across all BSs is equal to the sum of the weights (numbers that show clients' priorities) across all clients [Theorem 3]. Next, we use this property to prove that any equilibrium outcome of AFRA is {\it{globally optimal}} [Theorem~4]. Finally, we show that at equilibrium the vector of throughput rates across all clients is unique; however, there could be infinitely many resource allocations that realize this outcome [Theorem~5].

\item {\bf{Practicality:}} We construct a testbed with programmable BS hardware, and show that we can successfully aggregate the throughput across multiple BSs at the MAC layer. We also show that replacing the conventional resource allocation algorithm on each BS with AFRA can substantially increase the system throughput and fairness.

%
%

\item {\bf{Performance:}} We conduct extensive simulations to characterize AFRA's convergence time properties as we scale the number of BSs and clients. %
%
We also introduce policies
that reduce the convergence time by more than 30\%. Finally, we compare the performance of AFRA against DDNUM, a \underline{d}ual \underline{d}ecomposition algorithm that we derived from the \underline{NUM} framework. We show that compared to DDNUM, AFRA is 2-3 times faster with 4-5 times less over-the-air overhead. 

\end{itemize}

This paper is organized as follows. We discuss the related work in Section~\ref{sec:related}. We present the system model and details of AFRA 
in Section~\ref{sec:model}. In Sections~\ref{sec:convergence} and \ref{sec:optimality} we prove the convergence and optimality of AFRA. We present the results of our experiments, simulations, and comparisons against DDNUM in Section~\ref{sec:evaluation}. We conclude the paper in Section~\ref{sec:conclusion}. 




\section{Related Work}
\label{sec:related}
We discuss the related work in the areas of multi-BS communication and distributed optimization, and highlight their differences from this paper.

{\bf{Single-RAT Multi-BS Communication.}} Prior works have studied the problem of traffic aggregation when a client can simultaneously communicate with multiple same technology BSs. For example,~\cite{shakkottai} uses game theory to model selfish traffic splitting by each client in WLANs. On the other hand, the resource allocation problem in HetNets is primarily addressed at the BS side. Similarly,~\cite{Prasad2017} proposes an approximation algorithm to address the problem of client association and traffic splitting in LTE DC. 
Our algorithm (AFRA) goes beyond this and other related work by guaranteeing {\it{optimal}} resource allocation for {\it{any number of RATs and BSs}}. 
Other works have developed centralized client association algorithms to achieve max-min~\cite{bejerano2007} and proportional fairness~\cite{yang2008} in multi-rate WLANs. 
In contrast, the problem of resource allocation in HetNets needs to be solved in a fully distributed manner.


{\bf{Multi-RAT Communication.}} Resource allocation algorithms that realize the capacity gains in HetNets are still in their early stages. The problem of PF resource allocation for LWA was studied in~\cite{sarabjotPF}. In the proposed setup, each client has one LTE and one WiFi RAT. 
Further, there is only a single LTE BS in the network, and each client's throughput across its WiFi RAT is {\it{fixed}}. Next, the authors propose a water-filling based resource allocation algorithm at the LTE BS that achieves PF. Similarly, we show that the optimal PF resource allocation in a single BS can be interpreted as a form of water-filling. However, we use the observation to design an optimal algorithm for the generic problem with any number of BSs and client RATs, and explicitly model the impact of system dynamics 
on the throughput that each client gets from every BS. In our prior work~\cite{ICDCS2017}, we addressed the problem of max-min fair resource allocation in HetNets. 
However, even with opportunistic centralized network supervision over autonomous resource allocation at each BS we could not optimally solve the problem. Here, we focus on the PF objective, which is commonly implemented in BSs, and show that we can optimally solve the problem in a purely distributed manner. Other works have built testbeds to evaluate the over-the-air performance of MAC-level cross-RAT throughput aggregation~\cite{IMPL1, IMPL2,IMPL3,IMPL4}. All these works have relied on conventional scheduling algorithms on each BS and focused on higher layer transport and application performance. We experimentally show that replacing the conventional resource allocation algorithms with AFRA can substantially increase the system throughput and fairness.


{\bf{Distributed Network Utility Maximization (NUM).}} There is a large body of general results on the mathematics of distributed computation, some of which are summarized in standard textbooks such as~\cite{DISBOOK1, DISBOOK2}. 
More recently, the framework of NUM~\cite{MUNGNUM, KELLY, SHROFFNUM} has emerged as a mathematical tool to optimize layered network architectures. The framework allows for decomposition of a global optimization problem into subsets of local problems that are carried out distributedly and implicitly solve the global NUM problem. We have derived an alternative distributed algorithm (named DDNUM) by leveraging dual decomposition and the NUM framework. We will show through simulations that DDNUM is  2-3 times slower than AFRA (in terms of convergence time) and increases the over-the-air signaling overhead by 4-5 times. These disadvantages, coupled with the increased client side computational complexity and lack of compatibility with conventional BSs, make 
NUM-based algorithms impractical for multi-RAT traffic aggregation.



\section{System Model}
\label{sec:model}

We discuss the system model and the resource allocation algorithm that is autonomously executed by each BS. 

\subsection{Network Model}

\begin{figure}[htbp]
\centering
\includegraphics[width=0.87\columnwidth]{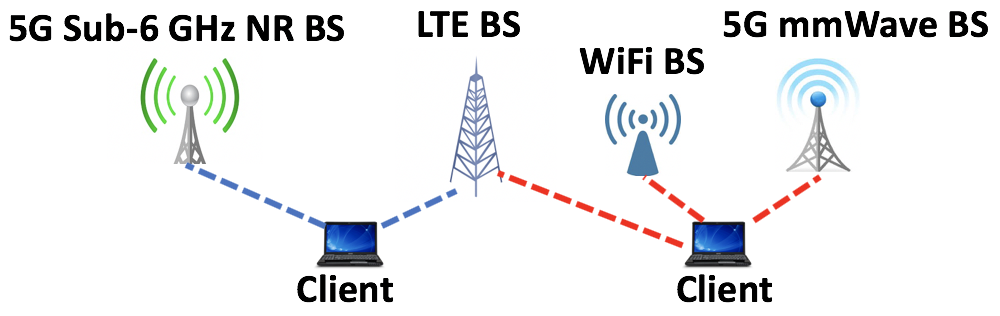}
\vspace{-0.1in}
\caption{{\bf{A heterogeneous network with 4 access technologies. Each client is in the coverage area of a group of BSs (dotted lines) and can split or aggregate its traffic across
the corresponding BSs (RATs). The 3GPP is actively developing several new RATs for both sub-6 GHz and mmWave bands, re-emphasizing the heterogeneity of future wireless networks. 
}}}
\label{fig:topology}
\end{figure}

We consider a HetNet composed of a set of BSs
$\M$ = $\{ 1, ..., M\}$ and a set of clients $\N$ = $\{1, ..., N\}$. 
Each BS has a limited transmission range
and can only serve clients within its range. Each client has
a client-specific number of RATs, and therefore has access to a subset of BSs. We model clients that can aggregate traffic across BSs of the same technology (\eg, LTE DC) with multiple such RATs. Fig~\ref{fig:topology} shows an example HetNet topology. 
We assume that clients split their traffic over 
the BSs 
and focus on the resource allocation problem at each BS. 
%
It is itself a challenging problem to determine which BS to associate with among same technology BSs (\eg, choosing the optimal LTE BS if a
client has an LTE RAT). We assume there exists a rule to pre-determine client RAT to BS association. The pre-determination rule
could for instance be any load balancing algorithm \cite{integration2011,andrews2013}, or based on the received signal strength. Similar to \cite{shakkottai, bejerano2007, yang2008, ICDCS2017, integration2011}, we assume that the transmission in one BS does not interfere with an adjacent BS. This can be achieved through spectrum separation between BSs that belong to different access networks and frequency reuse among same technology BSs. 

\subsection{Throughput Model}

We consider a multi-rate system and
use $R_{i,j}$ to denote the PHY rate
of client $i$ to BS $j$. Since each BS generally serves more than one client, clients of the same BS need to share resources such as time and
frequency slots (\eg in 3/4/5G) or transmission opportunities (\eg in WiFi). The throughput
achieved by client $i$ from BS $j$ thus depends on the load
of the BS and will be a fraction of $R_{i,j}$. We assume that each BS employs a TDMA throughput
sharing model\footnote{In Section~\ref{sec:implementation}, we discuss how we can extend our model and algorithm
to capture practical implementation issues such as WiFi contention.} and let $\lambda_{i,j}$ denote the fraction of time allocated to client $i$ by
BS $j$. Hence, the throughput achieved by client $i$ from BS $j$ is equal to $\lambda_{i,j}R_{i,j}$ and its total throughput across all its RATs would be
\vspace{-0.1in}
\begin{equation}
\label{eq:throughputmodel}
\textrm{Total Throughput of Client }i = r_i = \sum_{j=1}^{M}\lambda_{i,j}R_{i,j}
\end{equation}

The total amount of time fractions available to each BS cannot exceed 1. Thus, for the $\lambda_{i,j}s$ to be feasible we have

\begin{equation}
\label{eq:feasiblity1}
\sum_{i=1}^N\lambda_{i,j} \leq 1 \hspace{0.1in} \forall j \in \M
\end{equation}
\vspace{-0.1in}
\begin{equation}
\label{eq:feasiblity2}
\lambda_{i,j} \geq 0 \hspace{0.1in} \forall i \in \N, \hspace{0.05in} j \in \M
\end{equation}

\begin{table}[thbp]\caption{Main Notation}\label{table:tablen}
\centering
\small
\begin{tabular}{l }
$\mathbf{N}$ and $N$: Set and number of all clients in the network\\
$\mathbf{M}$ and $M$: Set and number of all BSs in the network\\
$R_{i,j}$: PHY rate of client $i$ to BS $j$ \\
$R_{max}$: maximum PHY rate across all clients and BSs \\
$R_{min}$: non-zero minimum PHY rate across all clients and BSs \\
$\lambda_{i,j}$: Fraction of time allocated to client $i$ by BS $j$ \\
$\blambda$: Vector of $\lambda_{i,j}$s across all clients and BSs\\
$r_i$: Total throughput of client $i$ across all its RATs\\
$\w_i$: A positive number that represents client $i$'s weight or priority\\
$\ttheta_j$: Water-fill level at BS $j$\\
\end{tabular}
\label{tab:TableOfNotationForMyResearch}
\vspace{-0.15in}
\end{table}

\subsection{Background: Conventional PF Allocation in a Single BS}
\label{sec:single-cell}

We first describe the basics of the PF resource allocation that is conventionally implemented in today's BSs. {\bf{Consider a network topology consisting of only a single BS {\it{j}} and $n'$ clients.}} Let $r_i$ denote the throughput of client $i$ and $\w_i$ a positive number that denotes its weight (or priority). A widely used objective function for PF is to maximize $\sum_{i=1}^{n'}\w_i\log(r_i)$~\cite{propfair1, propfair2}. It represents a tradeoff between throughput and fairness among the clients. Let $\lambda_i$ denote the time fraction allocated to client $i$ by BS $j$. To maximize the PF objective function, the BS needs to solve the following problem

\vspace{-0.1in}
\begin{eqnarray}
&\mathcal{P}_1: \hspace{0.1cm}  \max& \sum_{i=1}^{n'} \w_i\log(R_{i,j}\lambda_{i})  \nonumber\\
&s.t.&\sum_{i=1}^{n'}\lambda_{i} \leq 1 \hspace{0.4in}   \nonumber\\
&{\text{variables:}}&\lambda_{i} \geq 0 \hspace{0.37in}  \nonumber\\
\nonumber
\end{eqnarray}
\vspace{-0.3in}

Problem $\mathcal{P}_1$ can be easily solved through a simple algorithm. The Lagrangian of $\mathcal{P}_1$ can be expressed as 
\begin{equation}
L(\blambda, \mu) = \sum_{i=1}^{n'}\w_i\log(R_{i,j}\lambda_{i}) + \mu(1-\sum_{i=1}^{n'}\lambda_i)
\end{equation}

where $\mu$ is a constant number (Lagrange multiplier) chosen to meet the time resource constraint. Differentiating with respect to time fraction resource $\lambda_i$ and setting to zero gives

\begin{equation}
\label{eq:waterfill1}
\frac{R_{i,j}\w_i}{R_{i,j}\lambda_i} - \mu = 0 \implies \frac{\w_i}{\lambda_i} = \mu \hspace{0.1in} \forall i \in \{1,...,n'\}
\end{equation}

Since the sum of time fractions at optimality is equal to 1, we can conclude from Eq.~(\ref{eq:waterfill1}) that $\mu = {\sum\w_i}$. With known $\mu$ and $\w_i$, we can derive $\lambda_i$s from Eq.~(\ref{eq:waterfill1}).

Now, let $\ttheta_j$ be defined as $\frac{1}{\mu}$. Leveraging Eq.~(\ref{eq:waterfill1}), we have 
\begin{equation}
\label{eq:waterfill}
\frac{\lambda_i}{\w_i} = \ttheta_j \hspace{0.05in} \forall i\in \{1,...,n'\} \implies \boxed{\frac{r_i}{\w_iR_{i,j}} = \ttheta_j \hspace{0.05in} \forall i\in \{1,...,n'\}}
\end{equation}

Eq.~(\ref{eq:waterfill}) has an interesting water-filling based interpretation: 
{\it{the time allocated to each client is such that the throughput of the client divided by its PHY rate times its weight is the same across all clients}}. We refer to this ratio (\ie, $\ttheta_j$) as the water-fill level of BS $j$. In the next section, we will turn this observation in a single BS into a distributed resource allocation algorithm in HetNets.

\subsection{Distributed Resource Allocation in HetNets}
\label{sec:algAFRA}

There are two approaches to designing a resource allocation algorithm for generic HetNets. One approach, as we show in the Appendix, is to extend the formulation in $\mathcal{P}_1$ to include multiple BSs and client RATs, and use dual decomposition to derive a distributed algorithm. This approach converges to the optimal solution; however, the Lagrange multipliers across BSs would no longer correspond to BSs' water-fill levels. 
The second approach is to directly generalize the water-filling interpretation 
to derive an alternative algorithm, which still converges to the optimal solution (Section~\ref{sec:optimality}) with far less overhead, convergence time, and complexity than the dual decomposition based algorithm (Section~\ref{sec:DDNUM}).


%

\begin{figure}[htbp]
\centering
\includegraphics[width=0.83\columnwidth]{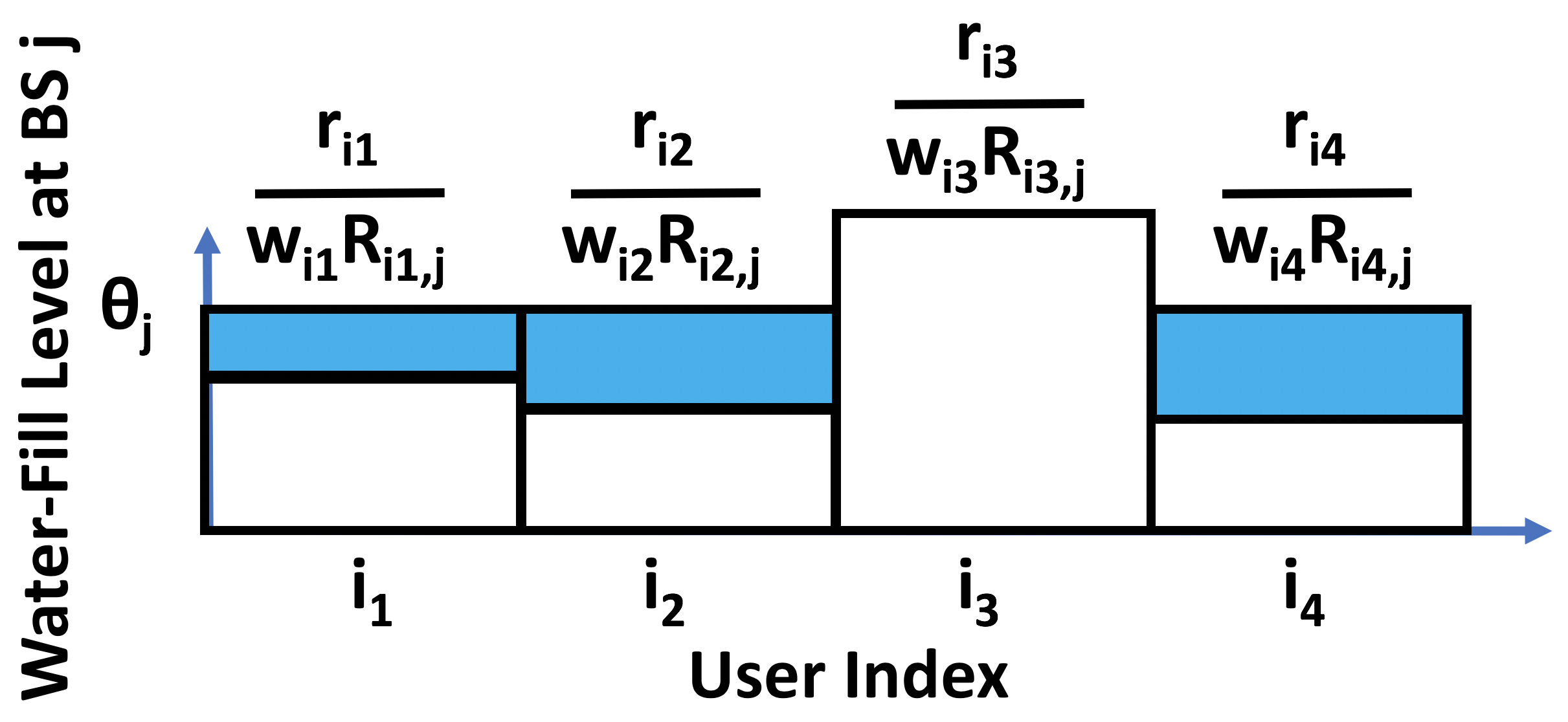}
\caption{{\bf{There are 4 clients with non-zero PHY rates to BS $j$. Blue boxes denote contributions to $\frac{r_i}{\w_iR_{i,j}}$ by BS $j$ (when it allocates time resources) and white boxes show contributions to it by other BSs. BS $j$ allocates its time resources so that all clients that get resources achieve the same water-fill level ($\ttheta_j$). Clients that do not get any resources from BS $j$ have a higher $\frac{r_i}{\w_iR_{i,j}}$ than $\ttheta_j$. Client $i_3$ is one such client in this example. 
}}}
\label{fig:waterfill}
\end{figure}

From Eq.~(\ref{eq:waterfill}), we observe that in a network with only a single BS, the BS allocates its time resources so that the clients who get the time resources reach the same water-fill level (\ie, throughput divided by $\w_iR_{i,j}$). Thus, in generic HetNets, if each BS considers the {\it{total throughput of each client across all its RATs}} (\ie, $r_i$) divided by $\w_iR_{i,j}$ in its water-fill definition, this should lead to a fair distributed algorithm. In other words, each BS $j$ should share its time resources across its clients such that: {\it{(1) all clients who get the time resources reach the same water-fill level at BS $j$ (\ie, $\ttheta_j$)
, and (2) if a client (\eg, $i'$) does not get any time resources from BS $j$, its $\frac{r_{i'}}{\w_{i'}R_{i',j}}$ is greater than $\ttheta_j$}}. Fig.~\ref{fig:waterfill} illustrates this operation.

\begin{figure}[htbp]
\centering
\includegraphics[width=1\columnwidth]{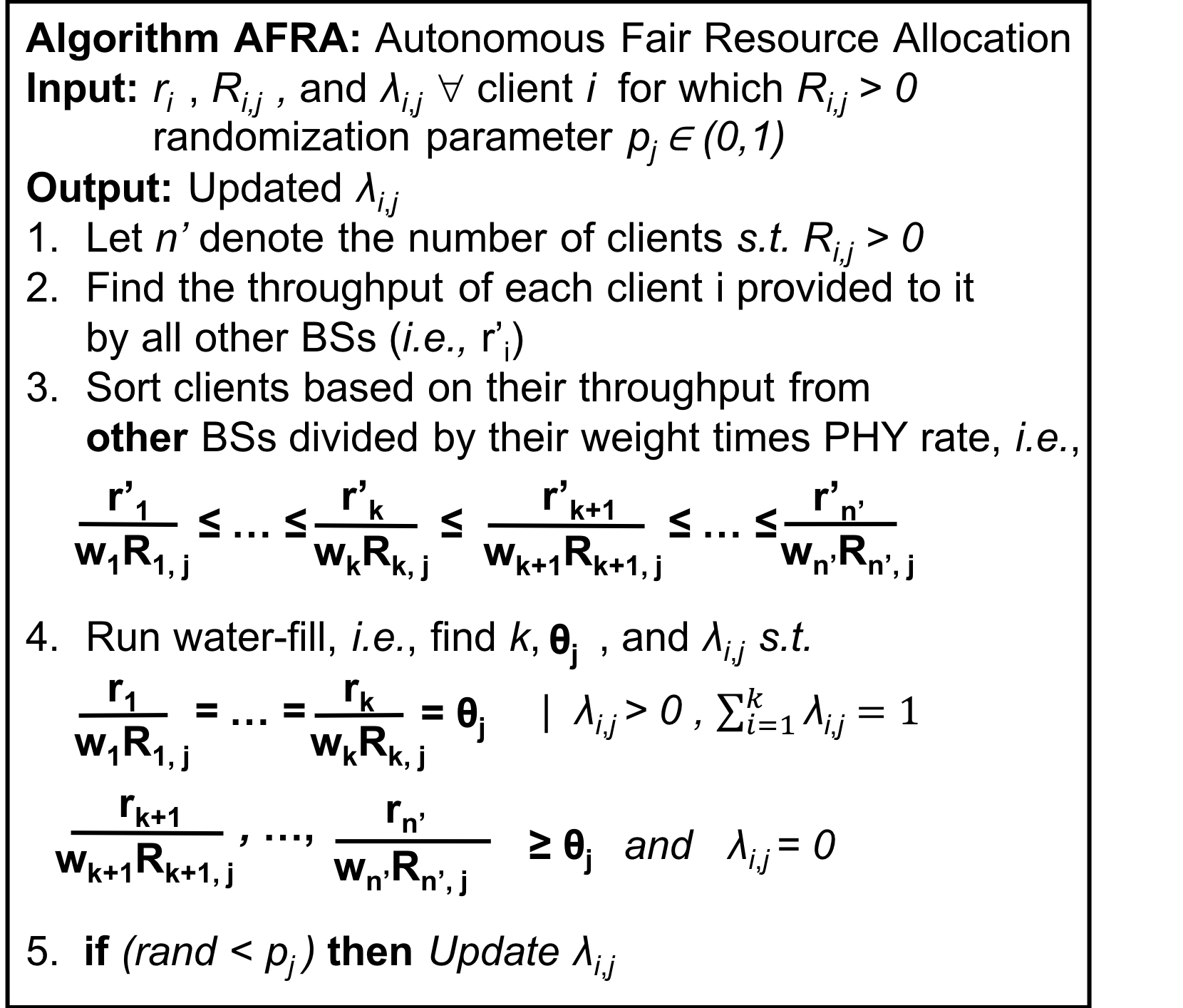}
\caption{{\bf{Resource allocation algorithm autonomously run by each BS $j$.
}}}
\label{fig:algorithm}
\vspace{-0.1in}
\end{figure}

We next turn this idea into a distributed resource allocation algorithm. Consider slotted time for now. Algorithm AFRA (Fig.~\ref{fig:algorithm}) summarizes the steps that are autonomously executed by each BS $j$. There are three main steps in the algorithm: (i) clients are sorted based on the total throughout they receive from other BSs ($r'_i$) divided by $\w_iR_{i,j}$ (Line 3), (ii) BS $j$ finds the water-fill level ($\ttheta_j$) and allocates the time resources accordingly (Line 4), and (iii) finally we introduce a randomization parameter to limit concurrent resource adaptation of a single client by multiple BSs (Line 5).  

We next elaborate on how each BS $j$ finds its water-fill level and its clients' time resource fractions (Line 4). 
Let $n'$ denote the number of clients such that $R_{i,j}>0$. Let $r'_i$ denote the total throughput of client $i$ from all BSs other than $j$. Consider an ordering in clients' $\frac{r'_i}{\w_iR_{i,j}}$ according to Line 3 of AFRA. In order to solve the water-fill problem (\ie, Line 4 of AFRA), we need to find the water-fill level $\ttheta_j$, client index $k$, and time fractions $\lambda_{i,j}$s such that 
\begin{gather}
\label{eq:walg1}
\frac{r'_1 + \lambda_{1,j}R_{1,j}}{\w_1R_{1,j}} = \frac{r'_2 + \lambda_{2,j}R_{2,j}}{\w_2R_{2,j}} = ... = \frac{r'_k + \lambda_{k,j}R_{k,j}}{\w_kR_{k,j}} = \ttheta_j\\
\label{eq:walg2}
\frac{r'_k}{\w_kR_{k,j}} < \ttheta_j \leq \frac{r'_{k+1}}{\w_{k+1}R_{k+1,j}}\\
\label{eq:walg3}
\sum_{i=1}^{k}\lambda_{i,j} = 1, \hspace{0.1in} \lambda_{i,j} >0
\end{gather}

We can find these variables with a simple set of linear operations. First, we can find $k$ by checking a set of inequalities

\begin{gather*}
\begin{cases}
\frac{\frac{r'_2\w_1R_{1,j}}{\w_2R_{2,j}}-r'_1}{R_{1,j}} \geq 1  \Rightarrow k = 1 \hspace{0.1in} \text{{\bf{else}}}\\
\frac{\frac{r'_3\w_1R_{1,j}}{\w_3R_{3,j}}-r'_1}{R_{1,j}} + \frac{\frac{r'_3\w_2R_{2,j}}{\w_3R_{3,j}}-r'_2}{R_{2,j}}\geq 1  \Rightarrow k = 2 \hspace{0.1in} \text{{\bf{else}}}\\
...\\
\frac{\frac{r'_{n'}\w_1R_{1,j}}{\w_{n'}R_{n',j}}-r'_1}{R_{1,j}} + ... + \frac{\frac{r'_{n'}\w_{n'-1}R_{n'-1,j}}{\w_{n'}R_{n',j}}-r'_{n'-1}}{R_{n'-1,j}}\geq 1\\
\hspace{2in}  \Rightarrow k = n'-1 \hspace{0.1in} \text{{\bf{else}}}\\
k = n'
  \end{cases}
  \end{gather*}
  
In the first inequality, we first check if $\frac{r'_1+R_{1,j}}{\w_1R_{1,j}} \leq \frac{r'_2}{\w_2R_{2,j}}$. If this is true, from Eq.~(\ref{eq:walg1}) we conclude that client 2 would have a higher $\frac{r'_2}{\w_2R_{2,j}}$ than $\frac{r_1}{\w_1R_{1,j}}$ even if BS $j$ allocated {\bf{all}} its time resources to client 1 (\ie, to the client with minimum $\frac{r'_i}{\w_iR_{i,j}}$ across all $n'$ clients). As a result $k$ should be equal to 1. This procedure (and logic) is continued until $k$ is found.

With known $k$, we can find $\ttheta_j$ by combining Eqs.~(\ref{eq:walg1}) and (\ref{eq:walg3}) and solving the following linear equation

\begin{equation}
\sum_{i=1}^{k}\frac{\ttheta_j\w_iR_{i,j} - r'_i}{R_{i,j}} = 1
\end{equation}

With known $k$ and $\ttheta_j$, the $\lambda_{i,j}$s can be found from Eq.~(\ref{eq:walg1}).

{\bf{AFRA's Computational Complexity and Message Passing Overhead.}} We calculate AFRA's computational complexity in finding the new time resource fractions ($\lambda_{i,j}$s) for a BS $j$. Let $n'$ denote the number of clients with non-zero PHY rates to $j$. The complexity of sorting clients (Line 3) is $O(n'\log(n'))$. The complexity of finding the water-fill level and the new time resource fractions (Line 4) is $O(n'\log(n'))$ (with a binary search to find $k$). Thus, the overall computational complexity is $O(n'\log(n'))$. If we assume that each client has on average $K$ RATs, then on average $n'$ would be equal to $\frac{KN}{M}$. Thus, the computational complexity would also be equal to $O(\frac{KN}{M}\log(\frac{KN}{M}))$.

Each BS uses the {\it{total throughput of each client across all its RATs}} in its calculations to find the water-fill level and the new $\lambda_{i,j}$s. Each time a client's time resource (and hence total throughput) is changed, the client needs to inform all BSs to which it is connected 
about its new total throughput. Thus, the total message passing overhead generated by clients of a single BS is at most equal to $O(n'K)$, or alternatively $O(\frac{K^2N}{M})$.


\section{Convergence and Speed of AFRA}
\label{sec:convergence}

In this section, we investigate the convergence properties of AFRA. We first show that as BSs autonomously execute AFRA, the system converges to an equilibrium. 
Next, we investigate the convergence time properties of AFRA and provide tight bounds to quantify it. 

\subsection{Convergence to an Equilibrium}

Before we discuss convergence, we present a formal definition of an equilibrium.
\begin{definition} Equilibrium: The vector of time fractions across all the BSs and clients is an equilibrium outcome if none of the BSs can increase its water-fill level through unilateral change of its time resource allocations.
\end{definition}

Our next theorem guarantees the convergence of AFRA.

\begin{frm-thm}
\label{theo:conv-AFRA}
\vspace{-0.1in}
Let each BS autonomously execute AFRA. Then, the system converges to an equilibrium, \ie, $\forall i   \in \N$ and $j\in\M$ $\lambda_{i,j}$ $\to$ $\lambda_{i,j}^{eq}$, $\theta_j$ $\to$ $\theta_{j}^{eq}$, and $r_i$ $\to$ $r_i^{eq}$.
\vspace{-0.1in}
\end{frm-thm}

\begin{proof}
Let $\blambda$ denote the vector of time fractions ($\lambda_{i,j}$s) across all clients and BSs, and 
$f(\blambda) = \sum_{i=1}^{N}\w_i\log(r_i)$ be the potential function. A potential function~\cite{potential} is a useful tool to analyze equilibrium properties, as it maps the payoff (\eg, throughput) of all clients into a single function.  
%
%


Since the number of clients and BSs is finite, $f$ is bounded. The key step to prove convergence, is to show that each time a BS $j$ adjusts its time fractions (\ie, $\lambda_{i,j}$s), 
the potential function ($f$) increases. This property coupled with $f$'s boundedness guarantees its convergence. We will show later in Eq.~(\ref{eq:provelimit}) that the change in potential function is proportional to the product of the change in water-fill levels and the change in $\lambda_{i,j}$s. Since $f$ converges (\ie, its variations converge to 0), one or both of these terms should converge to 0. Either of these conditions guarantee the convergence of the $\lambda_{i,j}$s (and hence, $\theta_j$s and~$r_i$s).


Next, we show that each time a BS runs AFRA, $f$ increases. When a BS runs AFRA, it takes some time resources from clients with high $\frac{r_i}{\w_iR_{i,j}}$ and distributes them across clients with lower values. To ease the proof presentation, we focus on two clients and follow the changes on $f$ as the BS adjusts the $\lambda_{i,j}$s dedicated to these clients. 

Let, $i, i'$ denote two clients who are currently receiving time resources from BS $j$. Assume the following initial (old) order between these two clients
\begin{equation}
\label{eq:orderc1}
\frac{r_i}{\w_iR_{i,j}} < \frac{r_{i'}}{\w_{i'}R_{i',j}}
\end{equation}

Therefore, as BS $j$ executes AFRA it changes the time resources from $\lambda_i$ and $\lambda_{i'}$ to $\lambda_i + \delta$ and $\lambda_{i'} - \delta$, respectively. This, only changes the two corresponding terms in the potential function, \ie

\begin{equation}
\label{eq:convergence1}
\begin{split}
f(\blambda)^{new} - f(\blambda)^{old} = \w_i\log(r_i + \delta R_{i,j}) - \\
\w_i\log(r_i) + \w_{i'}\log(r_{i'} - \delta R_{i',j})- \w_{i'}\log(r_{i'})=\\
\w_i\log(1+\delta\frac{R_{i,j}}{r_i}) + \w_{i'}\log(1-\delta\frac{R_{i',j}}{r_{i'}})
\end{split}
\end{equation}

Let $g(\delta)$ denote the variation in potential function, \ie
\begin{equation}
g(\delta) = \w_i\log(1+\delta\frac{R_{i,j}}{r_i}) + \w_{i'}\log(1-\delta\frac{R_{i',j}}{r_{i'}})
\end{equation}

Thus, to prove convergence, we need to prove that $g(\delta)$ is always positive. We prove this by showing that first $g'(\delta) \geq 0$. This shows that $g(\delta)$ is always non-decreasing. Second, we show that $g(\delta)$ is positive for very small values of $\delta$. Now

\begin{equation}
\label{eq:convergence3}
\begin{split}
g'(\delta) = \w_i\frac{\frac{R_{i,j}}{r_i}}{1+\delta\frac{R_{i,j}}{r_i}} - \w_{i'}\frac{\frac{R_{i',j}}{r_{i'}}}{1-\delta\frac{R_{i',j}}{r_{i'}}}\\
= \frac{\w_i R_{i,j}}{r_i + \delta R_{i,j}} - \frac{\w_{i'} R_{i',j}}{r_{i'} - \delta R_{i',j}} = \frac{\w_iR_{i,j}}{r_i^{new}} - \frac{\w_{i'}R_{i',j}}{r_{i'}^{new}} \geq 0
\end{split}
\end{equation}

\noindent
Here $r_{i}^{new}$ and $r_{i'}^{new}$ are the new throughput values for clients $i$ and $i'$, respectively. It is clear that after BS $j$ adjusts the time resources, we still have $\frac{r^{new}_i}{\w_iR_{i,j}} \leq \frac{r^{new}_{i'}}{\w_{i'}R_{i',j}}$. This is because after BS $j$ reduces $\lambda_{i',j}$, $\frac{r_{i'}^{new}}{\w_{i'}R_{i',j}}$ would be either equal to the new water-fill level or higher than it (if $\lambda_{i',j} = 0$). On the other hand, $\frac{r_{i}^{new}}{\w_iR_{i,j}}$ would be equal to the new water-fill level.
%
%
As a result, the final term in Eq.~(\ref{eq:convergence3}) is non-negative. Finally, $g(\delta)$ is greater than zero for small values of $\delta$ because 

\begin{equation}
\label{eq:provelimit}
\begin{split}
g(\delta) \stackrel{\textrm{Taylor Approx}}{\approx} \w_i\delta\frac{R_{i,j}}{r_i} - \w_{i'}\delta\frac{R_{i',j}}{r_{i'}} = \\\delta(\frac{\w_iR_{i,j}}{r_i} - \frac{\w_{i'}R_{i',j}}{r_{i'}}) > 0
\end{split}
\end{equation}

The last term in the above equation is due to Eq.~(\ref{eq:orderc1}).
\end{proof}

\vspace{-0.02in}
\subsection{Convergence Time}

Before we can derive a bound on convergence time, we need to define a discretization factor on the time fractions (\ie, $\lambda_{i,j}$s). This technicality is due to the fact that $\lambda_{i,j}$s in our model are continuous variables, 
%
which can cause some BSs to continuously make infinitesimal adjustments to them. These adjustments converge to 0 as time goes to infinity.

In practice, operations always happen in discretized levels. For example, consider the following discretization policy:

{\definition{Discretization Policy: During water-fill calculation by a BS j in AFRA, the time fraction allocated to the client with minimum $\frac{r_i}{\w_iR_{i,j}}$ should increase by at least $\epsilon$. Otherwise, the BS would not update its time fractions.}}

Based on the above discretization policy, we can derive the following bound on the convergence time.

\begin{frm-thm}
\vspace{-0.1in}
Consider a HetNet with N clients and M BSs. Then, the number of steps that it takes for AFRA to converge is upper bounded by O($\frac{NM^2\log(MN)}{\epsilon^2}$).
\vspace{-0.1in}
\end{frm-thm}

\begin{proof} Let $f(\blambda) = \sum_{i=1}^{N}\w_i\log(r_i)$ be the potential function from the proof of Theorem~\ref{theo:conv-AFRA}. To compute a bound on the convergence time, we study the increments of $f$. The key step is to find a lower bound on $f$'s increments. Since $f$ increases whenever a BS makes adjustments to its $\lambda_{i,j}$s, the convergence time is then upper bounded by the difference between the maximum and minimum possible values of $f$ divided by the lower bound on $f$'s increments.

We take the following steps to find a lower bound on the potential function's increments. Let $\{i_1, i_2, ..., i_q\}$ denote the set of clients with non-zero PHY rates to BS $j$ and assume the following initial (old) order among the clients
\begin{equation}
\label{eqct:order}
\frac{r_{i_1}^{old}}{\w_{i_1}R_{i_1,j}} \leq \frac{r_{i_2}^{old}}{\w_{i_2}R_{i_2,j}} \leq ... \leq \frac{r_{i_q}^{old}}{\w_{i_q}R_{i_q,j}}
\end{equation}

When BS $j$ executes AFRA, it adjusts the time fractions in a way that increases the time resources allocated to client $i_1$. Let $\epsilon_{i_1}$ denote the increase in client $1$'s time resources and $r^{new}_{i_1} = r_{i_1}$ its new throughput. Let $\epsilon_{i_p}$ denote the change in client $i_p$'s ($i_p \in \{i_2, ..., i_q\}$) time resources and $r^{new}_{i_p}$ its new throughput.
Hence, we have

\begin{gather}
\label{eqct:1}
r^{new}_{i_1} = r_{i_1}, \hspace{0.1in} r^{old}_{i_1} = r_{i_1} - \epsilon_{i_1}R_{i_1,j}\\
\label{eqct:2}
r^{new}_{i_p} = r_{i_p}, \hspace{0.1in} r^{old}_{i_p} = r_{i_p} + \epsilon_{i_p}R_{i_p,j} \hspace{0.1in} \forall i_p \in \{i_2, ..., i_q\}\\
\label{eqct:3}
\epsilon_{i_1} = \epsilon_{i_2} + \epsilon_{i_3} + ... + \epsilon_{i_q} 
\end{gather}

However, even after BS $j$ adjusts its time resources, $i_1$ would still have the minimum $\frac{r_i}{\w_iR_{i,j}}$ across all clients. This is due to the water-fill based operation in AFRA. As a result

\begin{gather}
\label{eqct:4}
\frac{r_{i_1}}{\w_{i_1}R_{i_1,j}} \leq \frac{r_{i_p}}{\w_{i_p}R_{i_p,j}} \hspace{0.1in} \forall i_p \in \{i_2, ..., i_q\}\\
\label{eqct:5}
\implies \frac{R_{i_p,j}}{r_{i_p}} \leq \frac{\w_{i_1}}{\w_{i_p}}\frac{R_{i_1,j}}{r_{i_1}}
\end{gather}

Next, we find a lower bound on the potential function's increments
\begin{equation}
\label{eq:step1}
\begin{split}
f(\blambda)^{old} - f(\blambda)^{new} = \\
\w_{i_1}\log(1-\frac{\epsilon_{i_1}R_{i_1,j}}{r_{i_1}}) + \sum_{p=2}^{q}\w_{i_p}\log(1+\frac{\epsilon_{i_p}R_{i_p,j}}{r_{i_p}}) \stackrel{\textrm{Eq.~(\ref{eqct:5})}}{\leq} \\
\w_{i_1}\log(1-\frac{\epsilon_{i_1}R_{i_1,j}}{r_{i_1}}) + \sum_{p=2}^{q}\w_{i_p}\log(1+\frac{\epsilon_{i_p}R_{i_1,j}}{r_{i_1}}\frac{\w_{i_1}}{\w_{i_p}}) 
\end{split}
\end{equation}

Let $W = \sum_{p=2}^{q}\w_{i_p}$ and $x_p = \frac{\epsilon_{i_p}R_{i_1,j}\w_{i_1}}{r_{i_1}\w_{i_p}}$. Since the logarithm is a concave function, from Jensen's inequality~\cite{Jensen},

\begin{equation}
\label{eq:jenapp1}
\begin{split}
\sum_{p=2}^{q}\w_{i_p}\log(1+x_p) = W\sum_{p=2}^{q}\frac{\w_{i_p}}{W}\log(1+x_p)\leq \\
W\log(\sum_{p=2}^{q}(\frac{\w_{i_p}}{W}+\frac{\w_{i_p}}{W}x_p)) = W\log(1+\sum_{p=2}^{q}\frac{\w_{i_p}}{W}x_p)
\end{split}
\end{equation}

%

Leveraging Eq.~(\ref{eq:jenapp1}), we conclude that Eq.~(\ref{eq:step1}) is 

\begin{equation}
\label{eq:taylorapprox}
\begin{split}
\leq \w_{i_1}\log(1-\frac{\epsilon_{i_1}R_{i_1,j}}{r_{i_1}}) + W\log(1+\frac{\w_{i_1}}{W}\frac{R_{i_1,j}}{r_{i_1}}\overbrace{\sum_{p=2}^{q}\epsilon_{i_p}}^{=\epsilon_{i_1}})=\\
\w_{i_1}[\log(1-z) + \gamma \log(1+ \frac{z}{\gamma})] \stackrel{\textrm{Taylor Series}}{\leq} -\w_{i_1}\frac{z^2}{2}\\
\implies f(\blambda)^{new} - f(\blambda)^{old} \geq \w_{i_1}\frac{z^2}{2}
\end{split}
\end{equation}

where $z = \epsilon_{i_1}\frac{R_{i_1,j}}{r_{i1}}$ and $\gamma = \frac{W}{\w_{i_1}}$. Note that since we seek an upper bound on convergence time, we can choose a small enough $\epsilon_{i_1}$ so that $z, \frac{z}{\gamma} < 1$. 
These assumptions increase the upper bound but allow us to use the Taylor series in Eq.~(\ref{eq:taylorapprox}). 
If we let $R_{min}$ and $R_{max}$ denote the minimum and maximum PHY rates across all the clients and BSs, then we have


\begin{equation}
\textrm{Convergence Time} \leq \frac{\textrm{Max} f(\blambda)  - \textrm{Min} f(\blambda)}{\frac{1}{2}\w_{min}\epsilon^2(\frac{R_{i_1,j}}{r_{i_1}})^2}\nonumber
\end{equation}

\begin{equation}
\leq \frac{(\sum_{i=1}^{N}\w_i)(\log(r_{max}) - \log(r_{min}))}{\frac{1}{2}\w_{min}\epsilon^2(\frac{R_{min}}{MR_{max}})^2} \leq\nonumber
\end{equation}

\begin{equation}
\frac{(\sum\w_i)(\log(MR_{max}) - \log(\frac{\w_{min}}{\sum\w_i}R_{min}))}{\frac{1}{2}\w_{min}\epsilon^2(\frac{R_{min}}{MR_{max}})^2}\nonumber
\end{equation}

\begin{equation}
\equiv O(\frac{NM^2\log(MN)}{\epsilon^2})
\end{equation}
\end{proof}

\section{Optimality of AFRA}
\label{sec:optimality}

Beyond convergence, we study the optimality properties of AFRA's equilibria. We first derive some useful properties of the equilibria that we leverage for optimality analysis. Next, we prove that the equilibria also maximize the global proportional fair resource allocation problem across all the BSs, and hence are globally optimal. Finally we discuss the uniqueness of the equilibria and prove that while the equilibrium throughput vector across all the clients is unique, there could be infinitely many resource allocations that realize this outcome. For simplicity, we do not consider discretization in this section.

\begin{frm-thm}
\label{theo:properties}
\vspace{-0.1in}
Consider an equilibrium outcome of AFRA. Let $r^{eq}_i$ denote the throughput of client i, 
$\theta^{eq}_j$ the water-fill level of BS j, and $\lambda^{eq}_{i,j}$ the fraction of time allocated to client i by BS j. Then \\
{\circled{I}} $\frac{\w_iR_{i,j}}{r^{eq}_i} \leq \frac{1}{\theta^{eq}_j} \hspace{0.1in}\forall i \in \N, j \in \M$ \\ 
{\circled{II}} $\sum_{i=1}^{N}\lambda_{i,j}^{eq} = 1 \hspace{0.1in}\forall j \in \M$\\
{\circled{III}} $\sum_{i=1}^{N} \w_i = \sum_{j=1}^{M}\frac{1}{\theta^{eq}_j}$
\vspace{-0.1in}
\end{frm-thm}

\begin{proof}
Part 1. From the water-fill definition we have

\begin{gather}
\label{eqp1}
R_{i,j}, \lambda^{eq}_{i,j} > 0 \implies  \frac{r^{eq}_{i}}{\w_iR_{i,j}} = \theta^{eq}_j\\
\label{eqp2}
R_{i,j}>0, \lambda^{eq}_{i,j} = 0 \implies \frac{r^{eq}_i}{\w_iR_{i,j}} \geq \theta^{eq}_j
\end{gather}

Property {\circled{I}} follows from Eqs.~(\ref{eqp1}) and~(\ref{eqp2}).

Part 2. Every BS can always increase its water-fill level by distributing its unused time resources across its clients. The property follows, since at equilibrium the water-fill levels cannot be further increased.

Part 3. We leverage {\circled{I}} and {\circled{II}} to derive property {\circled{III}} as follows 

\begin{gather}
\sum_{i=1}^{N}\w_i = \sum_{i=1}^{N}\frac{\w_ir^{eq}_i}{r^{eq}_i} = \sum_{i=1}^{N} \sum_{j=1}^{M} \frac{\w_i\lambda^{eq}_{i,j}R_{i,j}}{r^{eq}_i} = \sum_{\lambda^{eq}_{i,j}>0}\frac{\w_i\lambda^{eq}_{i,j}R_{i,j}}{r^{eq}_i}\nonumber\\
\stackrel{\textrm{Eq.~(\ref{eqp1})}}{=} \sum_{\lambda^{eq}_{i,j}>0}\frac{\lambda^{eq}_{i,j}}{\theta^{eq}_j} 
\stackrel{\circled{II}}{= }\sum_{j=1}^{M}\frac{1}{\theta^{eq}_j}
\end{gather}

\end{proof}

We next show that any equilibrium outcome of AFRA is globally optimal, \ie, it maximizes the global PF resource allocation problem. 

\begin{frm-thm}
\label{theorem:max}
\vspace{-0.1in}
Consider an equilibrium outcome of AFRA. Then, the equilibrium outcome also maximizes the global PF resource allocation problem, \ie, it maximizes $\sum_{i=1}^{N}\w_{i}log$($r_i$) subject to the feasibility constraints in Eqs.~(\ref{eq:throughputmodel})-(\ref{eq:feasiblity2}).
\vspace{-0.1in}
\end{frm-thm}

\begin{proof} Let $r^{eq}_i$ and $\theta^{eq}_j$ denote the throughput of client $i$ and water-fill level of BS $j$ at an equilibrium, respectively. 

%

We prove that for any feasible selection of $\lambda_{i,j}$s (\ie, $\lambda_{i,j}$s that satisfy the feasibility conditions in Eqs.~(\ref{eq:feasiblity1}) and (\ref{eq:feasiblity2})) and the corresponding clients' throughput values (\ie, $r_i$s as defined in Eq.~(\ref{eq:throughputmodel})) we have
\begin{equation}
\label{eq:globalproof}
\sum_{i=1}^{N}\w_i\log(\thr_i) \leq \sum_{i=1}^{N}\w_i\log(r^{eq}_i)
\end{equation}

Define $\textrm{W} = \sum_{i=1}^N\w_i$. 
Eq.~(\ref{eq:globalproof}) can then be proved through the following inequalities 
by leveraging properties {\circled{I}} and {\circled{III}} from Theorem~\ref{theo:properties}:

\begin{gather}
\sum_{i=1}^{N}\w_i\log(\thr_i) - \sum_{i=1}^{N}\w_i\log(r^{eq}_i) = \sum_{i=1}^{N}\w_i\log(\frac{\thr_i}{r^{eq}_i}) = \nonumber\\
\textrm{W}\sum_{i=1}^{N}\frac{\w_i}{\textrm{W}}\log(\frac{\thr_i}{r^{eq}_i})  \stackrel{\textrm{Jensen Inequality}}{\leq}\textrm{W}\log(\sum_{i=1}^{N}(\frac{\w_i}{\textrm{W}} \times\frac{\thr_i}{r^{eq}_i}))=\nonumber\\
\textrm{W}\log(\frac{1}{{\textrm{W}}}\times\sum_{i=1}^{N}(\frac{\w_i\thr_i}{r^{eq}_i})) = \textrm{W}\log(\frac{1}{\textrm{W}}\times\sum_{i=1}^{N}\sum_{j=1}^M\frac{\w_i\lambda_{i,j}R_{i,j}}{r^{eq}_i})\nonumber\\
\stackrel{\circled{I}}{\leq}\textrm{W}\log(\frac{1}{{\textrm{W}}}\times\sum_{i=1}^{N}\sum_{j=1}^M\frac{\lambda_{i,j}}{\theta^{eq}_j})
\stackrel{\textrm{Eq.~(\ref{eq:feasiblity1})}}{\leq}
\textrm{W}\log(\frac{1}{ \textrm{W}}\times\sum_{j=1}^M\frac{1}{\theta^{eq}_j})\stackrel{\circled{III}}{=} \nonumber\\
\textrm{W}\log(\frac{1}{\textrm{W}}\times\sum_{i=1}^N\w_i)=0 \label{eq:maximum}
\end{gather}


\end{proof}

In our last theorem we prove that while the equilibrium throughput vector across all clients is unique, there could be infinitely many resource allocations that realize this outcome.

\begin{frm-thm}
\vspace{-0.1in}
Let $\bbr^{eq}$ = $(r_{1}^{eq}, ..., r_{N}^{eq})$ denote the vector of throughput rates across all clients at an equilibrium. Then, $\bbr^{eq}$ is unique. However, there could be infinitely many resource allocations across the BSs that realize $\bbr^{eq}$.
\vspace{-0.2in}
\end{frm-thm}

\begin{proof}
Part 1. We first prove that $\bbr^{eq}$ is unique. 
Let $\bbr^{eq}$ maximize the global proportional-fair resource allocation across all clients and assume 
$\bbr'^{eq}$ is a different equilibria. 
From Theorem~\ref{theorem:max}, we know that every other equilibrium should also maximize the global PF  resource allocation. This means that all inequalities in Eq.~(\ref{eq:maximum}) should be equalities for any equilibrium, including $\bbr'^{eq}$. Now, for the first inequality to be an equality (\ie, Jensen inequality of Eq.~(\ref{eq:maximum})), the following condition needs to be satisfied~\cite{Jensen}

\begin{equation}
\label{eq:condition1}
\frac{r^{eq}_1}{r'^{eq}_1} = \frac{r^{eq}_2}{r'^{eq}_2} = ... = \frac{r^{eq}_n}{r'^{eq}_n} \implies r'^{eq}_i = \alpha \hspace{0.02in} r^{eq}_i \hspace{0.1in} \forall i \in \N
\end{equation}

Further, since $\sum_{i=1}^{N}\w_i\log(r^{eq}_i) = \sum_{i=1}^{N}\w_i\log(r'^{eq}_i)$, we conclude that
\begin{equation}
r^{eq}_i = r'^{eq}_{i} \hspace{0.1in} \forall i \in \N
\end{equation} 


Part 2. To prove that there could be infinitely many resource allocations that realize $\bbr^{eq}$, we provide an example. Consider a topology with two BSs ($j_1$, $j_2$) and two clients ($i_1, i_2$). Let $R_{i_1,j} = 1 \hspace{0.1in} \forall j\in \M$, $R_{i_2,j} = 2 \hspace{0.1in} \forall j\in \M$, and $\w_{i_1} = \w_{i_2} = 2$. Then, $\sum\w_i\log(r_i)$ is maximized by the following time fractions for any $\alpha \in [0 \hspace{0.1in}1]$.


\begin{equation}
\begin{split}
\lambda_{i,j} = \alpha \hspace{0.2in} \textrm{for}\hspace{0.1in} i=i_1, j=j_1 \hspace{0.1in} \textrm{and} \hspace{0.1in} i=i_2, j=j_2\\
\lambda_{i,j} = 1 - \alpha \hspace{0.1in} \textrm{for} \hspace{0.1in}i=i_1, j=j_2 \hspace{0.1in} \textrm{and} \hspace{0.1in} i=i_2, j=j_1
\end{split}
\end{equation}

Here, irrespective of $\alpha$, $r_{i_1} = 1$ and $r_{i_2} = 2$. 
%
%
%
\end{proof}

\section{Performance Evaluation}

\begin{figure*}
\centering
    \includegraphics[width=2\columnwidth]{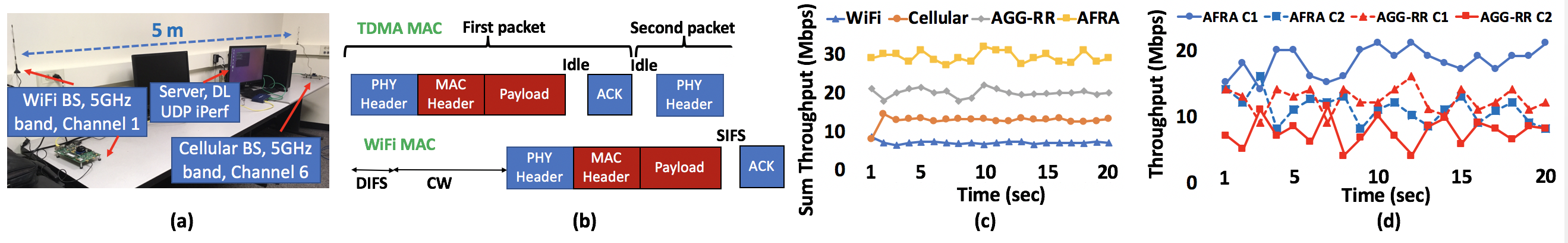}
\caption{{\bf{We use two WARP boards to construct two BSs in our testbed. The BSs are connected to a server through Ethernet. The server runs a single fully-backlogged DL UDP iPerf session to each client. A sublayer implementation below the IP layer at the server, selects the BS for each packet of every traffic flow. The clients (not shown in the photo) have access to both radios, and remain static and connected to both BSs throughput the experiments (a); Cellular TDMA and WiFi MACs. The PHY header and ACKs are sent at a fixed transmission rate. Clients embed the throughput they receive from other BS in their ACK packets. The MAC header and payload are transmitted at a variable transmission rate. We define $R^{eff}$ as the total number of payload bits divided by the total time it takes to successfully transmit a packet. We replace all $R_{i,j}$s in AFRA with $R^{eff}_{i,j}$ to derive the $\lambda_{i,j}$s and determine the number of packets that should be served from each queue (b); Total throughput across the two clients for four schemes: WiFi only (WiFi), Cellular only (Cellular), AGG-RR, and AFRA. AFRA achieves a higher average total throughput (29 Mbps vs 20 Mbps) and PF index (2.3 vs 1.97) compared to AGG-RR(c); Per-client throughput values for both AFRA and AGG-RR (d).}}}
\label{fig:testbed}
\end{figure*}

\label{sec:evaluation}

In this section, we evaluate AFRA's performance through experiments and simulations. First, we investigate the benefits of MAC level traffic aggregation in a small testbed composed of four SDR (software-defined radio)-based BSs and clients. Next, we conduct simulations to evaluate AFRA's equilibria properties as we scale the number of clients and BSs. Finally, we compare AFRA's 
%
speed and over-the-air signaling overhead against DDNUM, a \underline{d}ual \underline{d}ecomposition based algorithm that we derived from the \underline{NUM} framework. 

\subsection{SDR-Based Implementation and Real-World Performance}
\label{sec:implementation}

{\bf{Implementation.}} We construct a HetNet topology composed of a WiFi BS, a cellular BS, and two clients. The two BSs are physically separated from each other and are placed in an indoor lab environment (Fig.~\ref{fig:testbed}(a)). We use a WARP board~\cite{WARP} with 802.11a reference design as our WiFi BS. We use another WARP board with OFDM PHY (WARP OFDM reference design) and a custom TDMA (Time Division Multiple Access) MAC to mimic a cellular BS. 
We use two other WARP boards to construct our two clients. Each client has access to both WiFi and cellular radios, and remains static and connected to both BSs throughout the experiments. 

A server running iPerf sessions is connected to both BSs through Ethernet. For each client, the server generates a {\bf{single fully-backlogged UDP traffic flow}} with 500 byte packets. We implement a below-IP sublayer to split this traffic flow between the two BSs. This sublayer is responsible for selection of the BS to be used for each packet, and acts similar to the LWA Adaptation Protocol (LWAAP) in the LWA standard~\cite{3GPPLWA}. In our implementation, we sequentially iterate between the WiFi and cellular BSs to route the packets of each traffic flow.

AFRA, as presented in Section~\ref{sec:algAFRA}, does not account for various types of overhead (\eg, PHY/MAC header, ACKs, idle slots, collisions) that exist in PHY/MAC protocols. To address the issue, we introduce the notion of effective rate ($R^{\text{eff}}$) and replace all $R_{i,j}$s in AFRA with $R^{\text{eff}}_{i,j}$s. For a single packet, $R^{\text{eff}}$ can be calculated as the number of bits in the packet divided by the {\it{total time}} it takes by a BS to successfully transmit that packet (including all overhead). In our implementation, each BS keeps track of the total time spent in successfully transmitting the past 5 packets of each traffic flow (\ie, the past 5 packets of each client) to calculate its $R^{\text{eff}}_{i,j}$. The averaging over 5 packets is to account for channel fluctuations in our experiments, and can be adjusted based on the client mobility.

We implement the following mechanisms: (i) WiFi only: the cellular BS is off but the WiFi BS is active; (ii) Cellular only: WiFi BS is off; (iii) AGG-RR: this scheme uses aggregation but with a round robin (RR) scheduler at the WiFi BS and conventional PF MAC at the cellular BS. With the RR scheduler, the WiFi BS maintains a different queue for each client and sequentially serves a single packet from each queue at every round. With the PF MAC at the cellular BS, the BS dedicates its time resources to each client according to Section~\ref{sec:single-cell} (single BS PF); (iv) AFRA: each BS uses its calculated $\lambda_{i,j}$s to determine the number of packets that should be served from each queue in WiFi and the number of time slots that should be dedicated to each queue (client) in cellular, at every round. In our implementation, both clients' $\w_i$ are equal to 1 and the BSs updates their $\lambda_{i,j}$s every 5~ms.

\begin{figure*}[htp]
\centering
\mbox{
\subfigure[]
{
    \includegraphics[width=0.46\columnwidth]{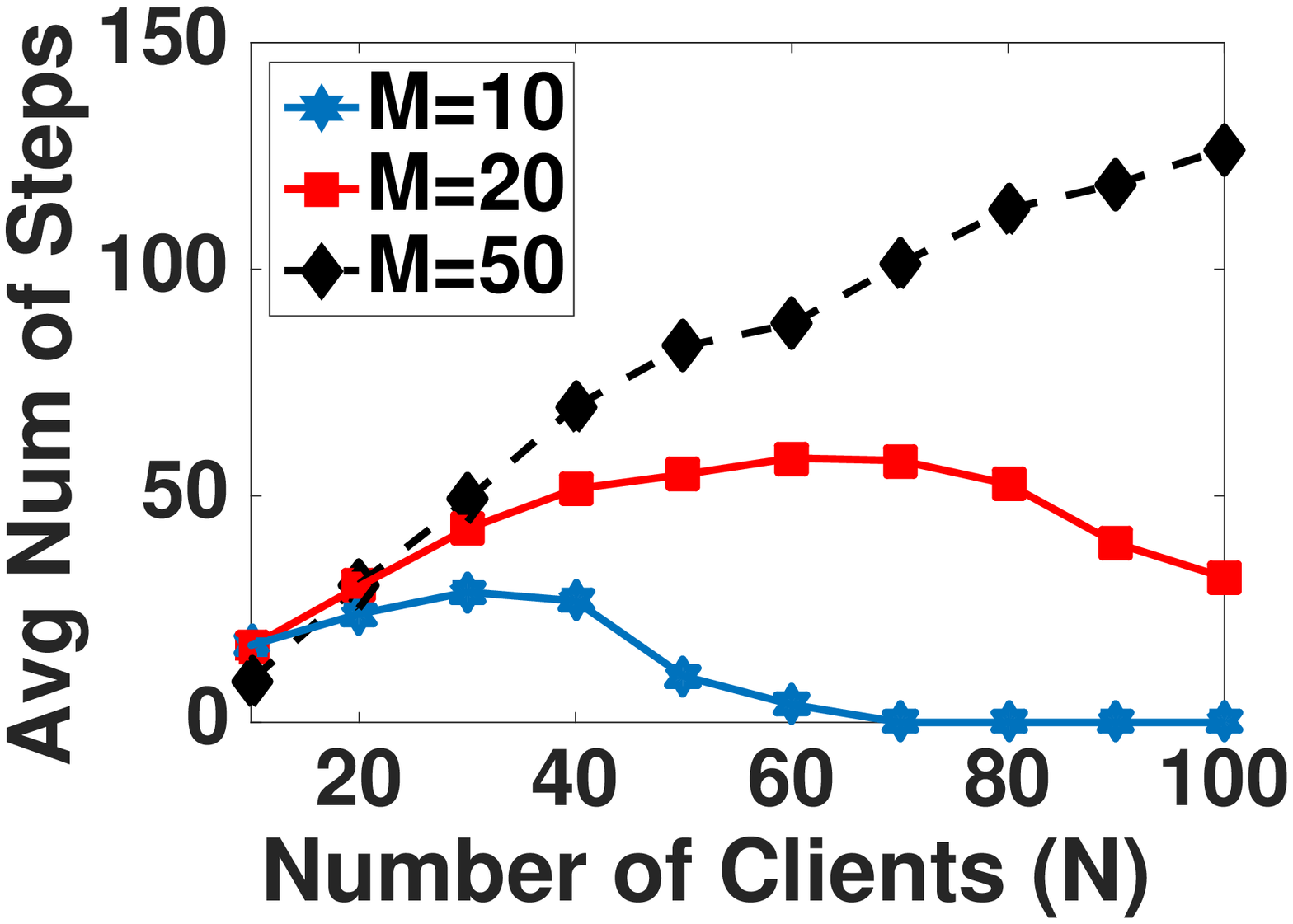}
}
\subfigure[]
{
    \includegraphics[width=0.46\columnwidth]{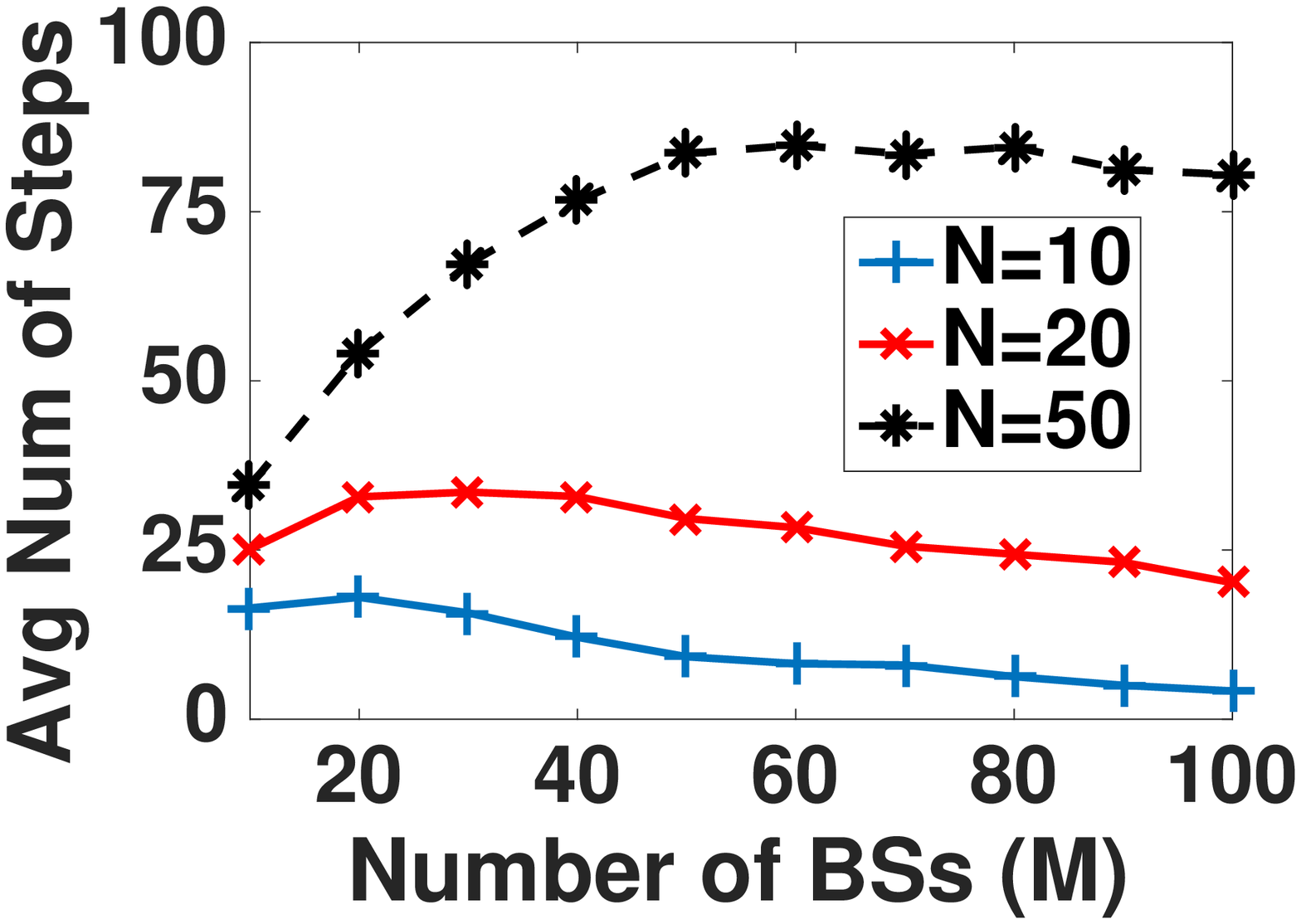}
}
\subfigure[]
{
    \includegraphics[width=0.46\columnwidth]{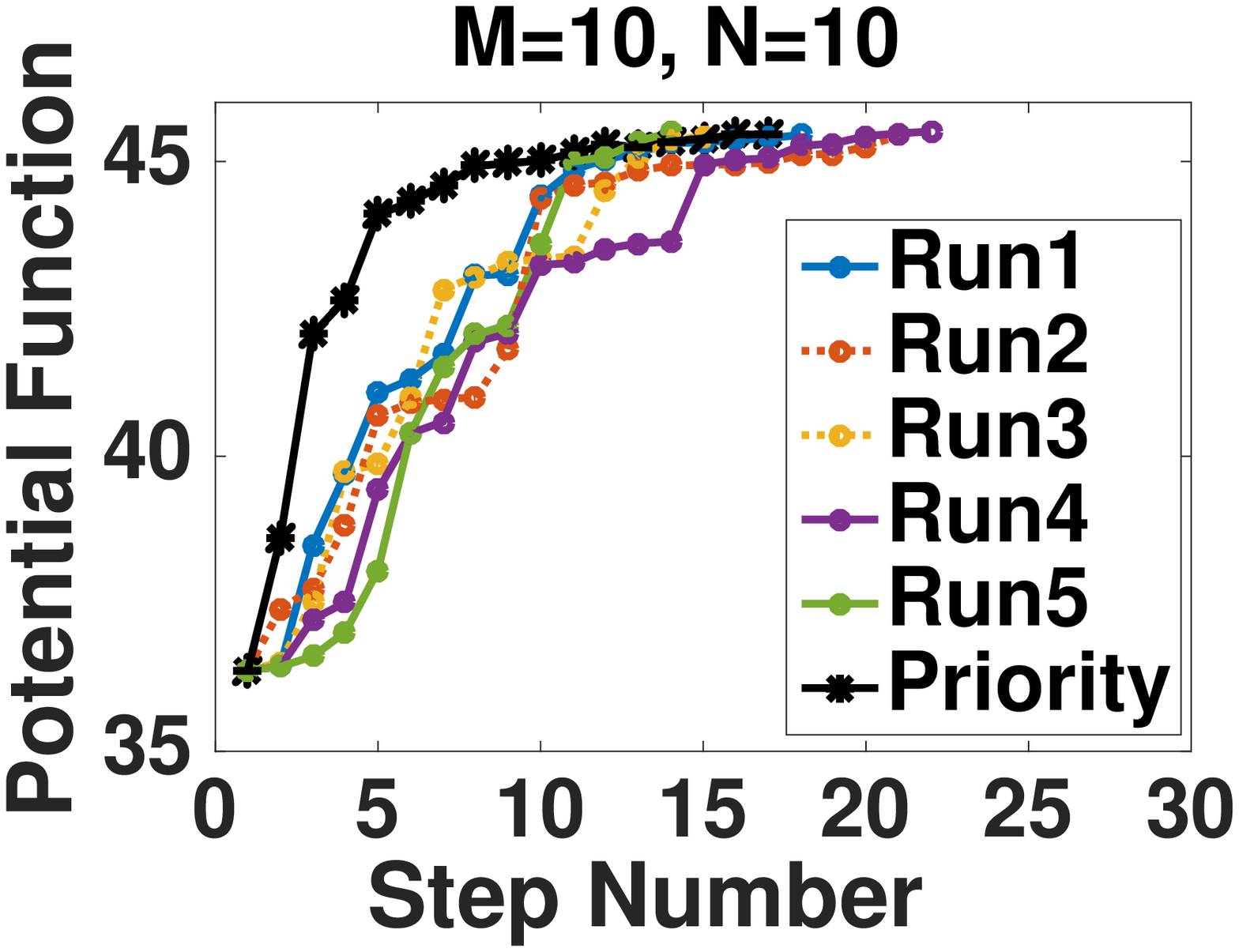}
}
}
\subfigure[]
{
    \includegraphics[width=0.46\columnwidth]{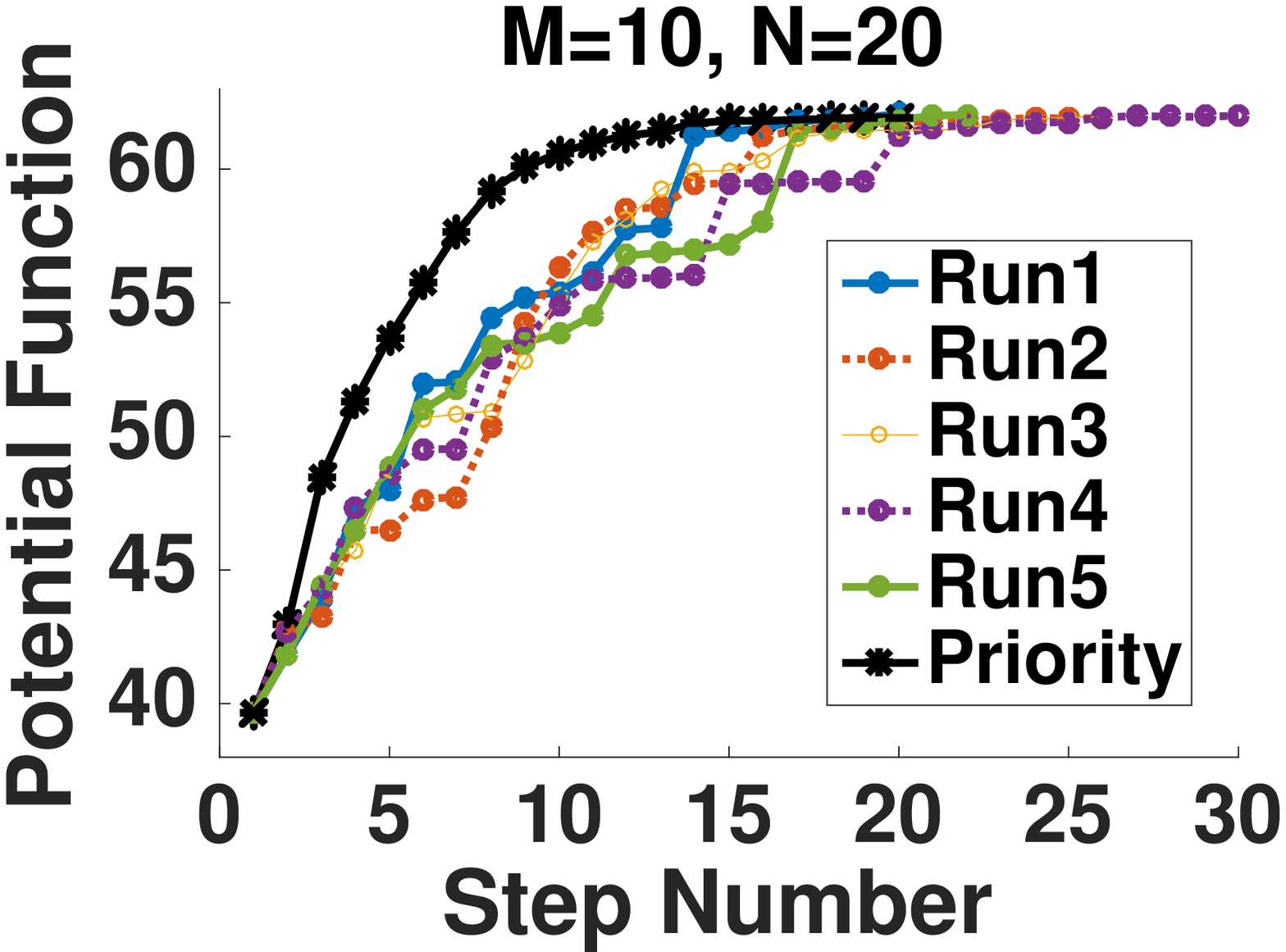}
}
\caption{{\bf{AFRA's performance evaluation results. Average number of steps to convergence as a function of number of clients (a) and number of BSs (b). Evolution of potential function for two simulation scenarios one with M=10, N=10 (c) and the other with M=10, N=20 (d). Each Run in these figures corresponds to a different simulation realization. In the priority curves (solid black curve with * markers), the BS with the highest local increase in potential function gets priority in executing AFRA. Leveraging this policy reduces the average convergence time by more than 30\%}.}}
\label{fig:evaluation}
\vspace{-0.2in}
\end{figure*}

{\bf{Performance Results.}} Fig.~\ref{fig:testbed}(c) shows the performance of the four schemes. In both the {\it{WiFi only}} and {\it{Cellular only}} options, only a single BS is active throughout the experiments. We observe that the {\it{Cellular only}} scheme provides a higher sum throughput than the {\it{WiFi only}} scheme. With careful evaluation of packet transmission traces, we discovered that this higher throughput is primarily due to the corresponding MAC protocols. In particular, WiFi MAC provides the same {\it{transmission opportunity}} to each traffic flow (client). As a result, the client with lower PHY rate occupies the channel for a longer duration that the other client. This decreases the throughput for both clients. In contrast, the cellular TDMA MAC provides the same {\it{transmission time}} for both clients (with 2 clients, single BS PF equally divides the time between the clients (Eq.~\ref{eq:waterfill1})). As a result, the throughput of the client with higher PHY rate does not drop because of the client with a lower PHY rate. This, along with other MAC issues such as WiFi contention reduce the WiFi only throughput.


Fig.~\ref{fig:testbed}(c) also shows that the two RAT aggregation schemes (AGG-RR and AFRA) can successfully aggregate WiFi and cellular capacities and provide a higher sum throughput than the {\it{WiFi only}} and {\it{Cellular only}} options. Further, AFRA increases the average total throughput by 45\% (from 20 to 29 Mbps) with 18 and 11 Mbps per-client total throughput values (per-client throughput plots are shown in Fig.~\ref{fig:testbed}(d)). 
%
Let us define the proportional fairness index as PF = $\sum_{i=1}^{2}\log(r_i)$ ($r_i$ is the total throughput of each client across its RATs in Mbps). Then, the PF index in AFRA would be 2.3. With AGG-RR, the per-client throughput rates  drop to 12.5 and 7.5 Mbps. Thus, the PF index reduces to 1.97. AGG-RR uses the conventional scheduling algorithms on each BS (\ie, it uses RR in WiFi and single BS PF in cellular), which reduce both the sum throughput and the PF fairness index.

\subsection{AFRA's Equilibria Properties}
\label{sec:AFRAeval}

{\bf{Setup.}} We simulated network deployments with N clients and M BSs to evaluate AFRA's equilibria properties as we scale the number of clients and BSs. All clients' $\w_i$s are equal to 1. Half of the BSs are WiFi and the other half are cellular. Each client has access to 4 RATs, two WiFi and two cellular. The PHY rates for the WiFi and cellular RATs are randomly selected from the sets $\{1, 2, 5.5, 11\}$ Mbps and $\{5.2, 10.3, 25.5, 51\}$ Mbps, respectively. In each simulation realization, we randomly associate clients' RATs with BSs. Next, we run AFRA until an equilibrium is reached. We set the discretization factor $\epsilon$ equal to 0.05, \ie, a BS adjusts its time fractions 
only if the increase in time fraction (\ie, $\lambda_{i,j}$) at its client with minimum $\frac{r_i}{\w_iR_{i,j}}$ is greater than or equal to 0.05. For the initial allocation, each BS equally divides its time across its clients. Unless otherwise specified, each of our simulation points is an average of 100 simulation realizations. 

{\bf{AFRA's Convergence Time.}} Figs.~\ref{fig:evaluation}(a) and~\ref{fig:evaluation}(b) depict the impact of the number of clients and BSs on AFRA's convergence time. In each of these figures, we count the number of steps until convergence is reached. At each step, a single BS that needs to adjust its time fractions 
is randomly selected. In Fig.~\ref{fig:evaluation}(a), we vary the number of clients from 10 to 100 and plot the corresponding convergence times for three different M values: 10, 20, and 50. We repeat this simulation by changing the N and M variables and plot the corresponding results in Fig.~\ref{fig:evaluation}(b). 
%
From these two figures, we observe that time to convergence is highest when the number of clients is between one to two times the number of BSs. As the ratio between the number of clients and BSs (\ie, $\frac{N}{M}$) 
leaves this range, the convergence time rapidly drops and then stabilizes. The results show that AFRA requires a small number of steps to reach an equilibrium.



{\bf{Policies to Further Reduce AFRA's Convergence Time.}} Our next goal is to design policies that can further reduce AFRA's convergence time. To gain intuition on how to design such policies, we simulated a topology with 10 clients and 10 BSs and plotted the evolution of the potential function (\ie, $\sum_{i}\log(r_i)$) 
as BSs adjusted their time fractions. The results are shown in Fig.~\ref{fig:evaluation}(c). Here, each Run corresponds to a different simulation realization. From these realizations we make two observations. First, there is a wide gap in the convergence times. Second, a high jump in the potential function pushes the system closer to equilibrium. Based on these observations, we designed a {\it{prioritization policy}} among the BSs 
to reduce the convergence time. 

We let each BS calculate the increase in the potential function assuming that it is the only BS executing AFRA.  
%
Since in AFRA each BS knows the current total throughput of its clients, it has all the needed information to calculate the increase in the potential function due to its action. 
Next, each BS broadcasts its calculated value. Finally, the BS with the highest value gets priority in executing AFRA. This distributed policy can be easily implemented in networks where all the BSs are connected to the same backbone (\eg, Ethernet). The solid black curve in Fig.~\ref{fig:evaluation}(c) shows the potential function's evolution with this policy. We observe that on average, the convergence time drops from 15 steps to 10, \ie, the prioritization policy reduces the convergence time by 33\%. We repeated this simulation for another setup with 20 clients to increase the topological redundancy. The results are plotted in Fig.~\ref{fig:evaluation}(d). Similarly, the average convergence time reduces from 19 steps to 13, \ie, a 32\% reduction in convergence time.

\subsection{Comparison Against DDNUM}
\label{sec:DDNUM}


We have compared AFRA's performance against DDNUM, a distributed algorithm that we developed by leveraging dual decomposition and the NUM framework. %
%
Dual decomposition is appropriate to solve the multi-RAT PF allocation problem, because the coupling constraint (Eq.~(\ref{eq:feasiblity1})) can be relaxed through the dual problem and then the problem decouples into subproblems that can be iteratively solved by clients and BSs. 

DDNUM is in essence similar to the standard dual algorithm presented in~\cite{MUNGNUM} to solve the basic NUM problem. We modified the algorithm in~\cite{MUNGNUM} to capture the constraints of our problem. 
At a high level, DDNUM has three main steps (for detailed algorithm derivation and discussions, refer to the Appendix):

\begin{itemize}
\compactlist

\item {\bf{Step 1:}} Initialization: set $t = 0$ and $\bmu(0)$ to some nonnegative value for each BS. Here, $\bmu(t)$ 
is the vector of Lagrange multipliers that shows the cost or congestion across all BSs. Each BS broadcasts its $\mu_j(0)$ to clients with $R_{i,j} > 0$.

\item {\bf{Step 2:}} Each client $i$ locally solves its Lagrangian problem, $\ie$, finds its time fractions ($\lambda^{*}_{i,j}(\mu_{j}(t))$) for each BS with $R_{i,j}>0$ and informs those BSs.


\item {\bf{Step 3:}} Each BS updates its price with a step size $\gamma$ and broadcasts the new price $\mu_{j}(t+1)$ to all its clients.

%
%

\end{itemize}

This procedure is repeated until a satisfying termination point is reached (\eg, the solution is within a desired proximity of the optimal solution). 
Similar to AFRA, DDNUM is guaranteed to converge and maximize the global optimization problem. 
However, there are several practicality and performance issues. We highlight a few of these issues next.

{\bf{Setup.}} To compare AFRA to DDNUM, we used the simulation setup in Section~\ref{sec:AFRAeval} (without the BS prioritization policy). We first run AFRA and let the system converge to an equilibrium. Next, we consider the 95\% value of AFRA's potential function at equilibrium as the desired algorithm termination point. We count the number of steps to reach the termination point and the resulting over-the-air signaling overhead in each of these two schemes. In DDNUM, the step size $\gamma$ (step 3) provides a balance between the final throughput values and speed. We choose the $\gamma$ that results in the fastest convergence time, subject to the potential function reaching the termination point. Finally, both AFRA and DDNUM can operate in either parallel or sequential mode with similar relative performance. We present the sequential mode results, \ie, at each time only a single BS adjusts its water-fill level (in AFRA) or announces a new price (in DDNUM). We assume that clients immediately update their BSs about their new throughput values (in AFRA) and desired $\lambda_{i,j}$s (in DDNUM) with no impact on the convergence time (similar to an FDD system in which uplink data is immediately available).

\begin{figure}[t]
\centering
\subfigure[]
{
    \includegraphics[width=0.46\columnwidth]{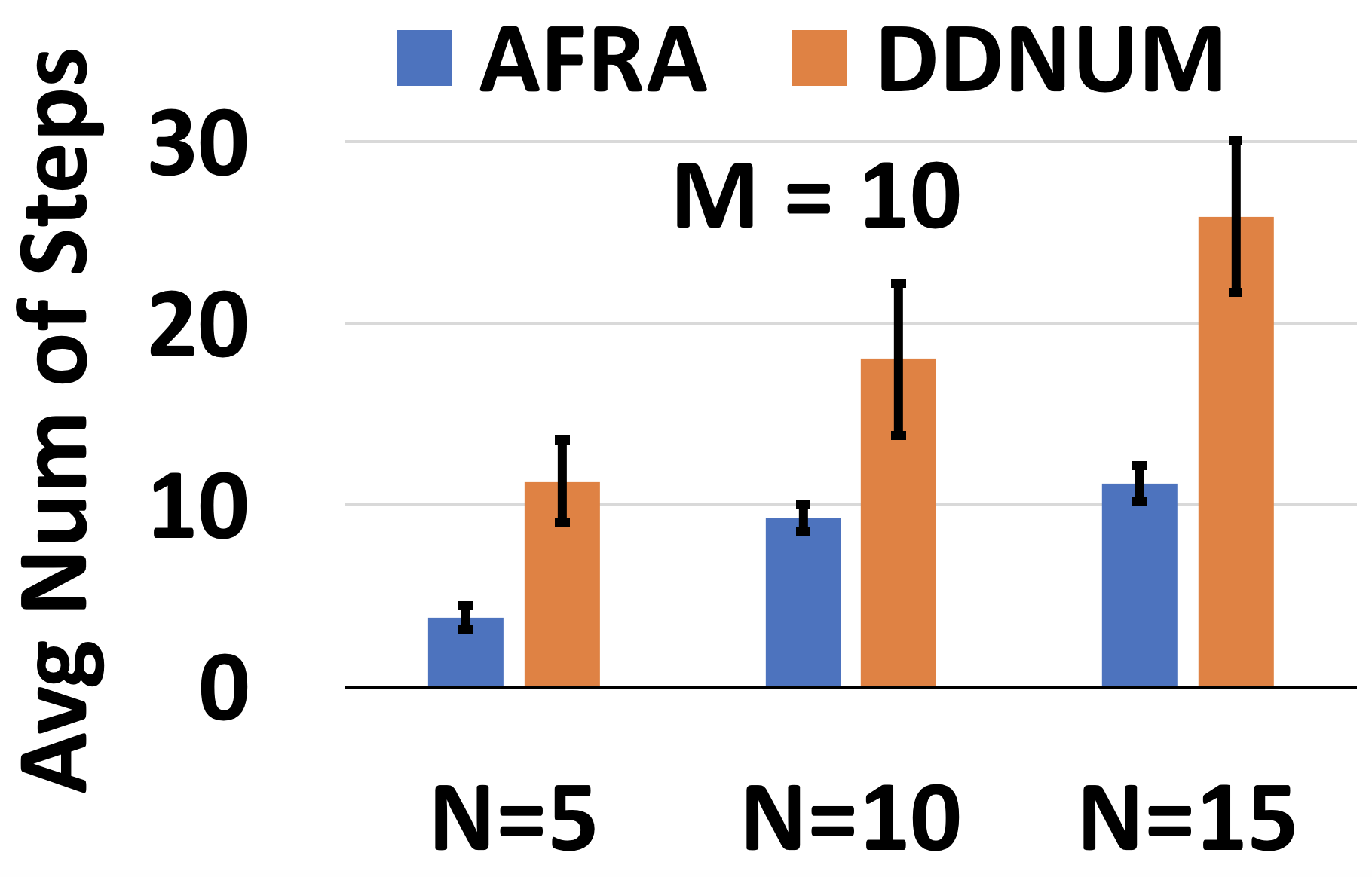}
}
\subfigure[]
{
    \includegraphics[width=0.46\columnwidth]{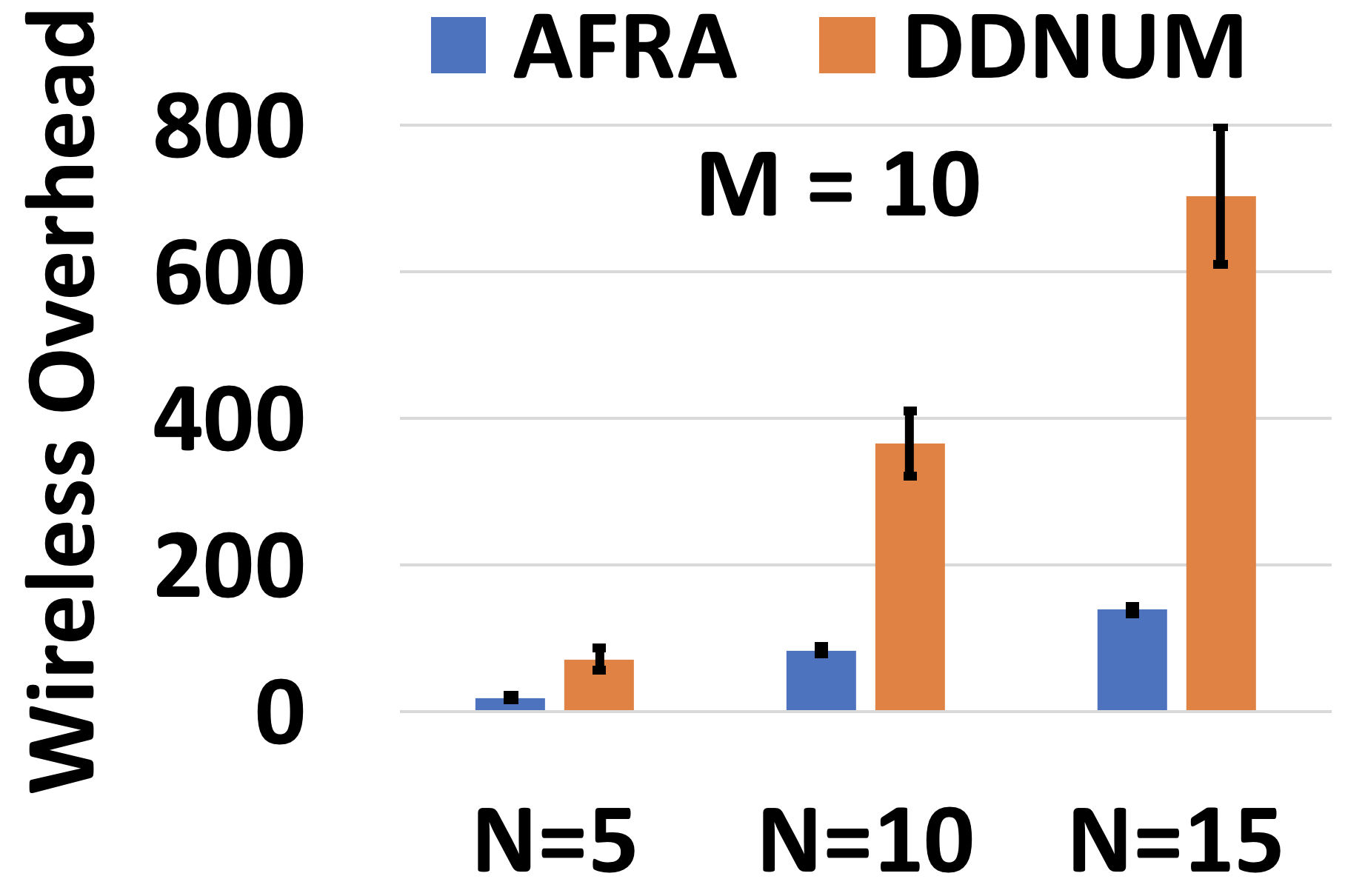}
}
\caption{{\bf{Compared to AFRA, DDNUM increases the average convergence time by 2.4x and the average over-the-air signaling overhead by 4.5x.}}}
\label{fig:DDNUM}
\vspace{-0.2in}
\end{figure}

{\bf{Speed.}} Fig.~\ref{fig:DDNUM}(a) show the convergence time results for a scenario with 10 BSs and varying number of clients. We observe that irrespective of the number of clients, DDNUM increase the convergence time by a factor of 2-3x with an average of 2.4x. In AFRA, each BS simultaneously calculates the water-fill level and finds the corresponding time fraction for each client. In DDNUM, the pricing mechanism requires a high number of iterations so that clients can find their optimal time fractions. This increases the convergence time.

{\bf{Over-the-Air Overhead.}} Fig.~\ref{fig:DDNUM}(b) shows the wireless signaling overhead results of the two schemes. We observe that DDNUM increases the signaling overhead by a factor of 4-5x with an average of 4.5x. There are several factors that contribute to DDNUM's high signaling overhead. First, the increases in convergence time results in a similar multiplicative increase in overhead. Second, in DDNUM both BSs and clients contribute to overhead. BSs continuously broadcast new prices and clients continuously inform each of their BSs about their desired time fractions. In contrast, in AFRA only clients update the BSs regarding their new throughput values. Third, with careful examination of simulation traces, we observed that in AFRA the water-fill operation only impacts a few of a BS's clients each time. In DDNUM, each time a BS updates its price, most of its clients would request new time fractions.

{\bf{Practicality.}} In DDNUM, each BS broadcasts its price while each client finds its desired $\lambda^*_{i,j}$s from its BSs. However, in real wireless systems BSs are responsible for resource allocation. Note that in DDNUM, it is not practical to shift the calculation of $\lambda^*_{i,j}$s (\ie, step 2) to BSs. This is because in order for a BS $j$ to find the $\lambda^*_{i,j}$s for each of its clients (\eg, $i$), it  would require knowledge about the client's $R_{i,j}$ and $\mu_j$ to every other BS for which the client's rate (\ie, $R_{i,j}$) is greater than zero. This information is only available at the client and pushing it to the BS would significantly increase the overhead, which is already very high in DDNUM.


{\bf{Complexity.}} In DDNUM, each client has to solve a complex Lagrangian subproblem to find its desired time fraction for each BS (step 2). This increases the computational complexity on the client devices. In contrast, AFRA identifies the time resources at the BSs, which have higher power and computing resources. Moreover, as we discussed in Section~\ref{sec:algAFRA}, AFRA has a very low total computational complexity.



\section{Conclusion}
\label{sec:conclusion}

We addressed the problem of proportional fair multi-RAT traffic aggregation in HetNets. We studied the conventional PF resource allocation in a single BS and showed that we can look at the problem as a special type of water-filling. 
Based on this observation, we designed a new fully distributed water-filling algorithm for HetNets. 
We also studied the convergence, speed, and optimality of our algorithm. We proved that our algorithm quickly converges to equilibria and derived tight bounds to quantify its speed. We also studied the characteristics of the optimal outcome, and used the properties to prove the outcomes of our algorithm are globally optimal.

\bibliographystyle{IEEEtran}
\bibliography{references}

\appendix \label{sec:appendix}

To maximize the PF objective function in generic multi-RAT HetNets we need to solve the following problem

\vspace{-0.1in}
\begin{eqnarray}
&\mathcal{P}_2: \hspace{0.1cm} \max &\sum_{i=1}^{N} \w_i\log(r_{i})  \nonumber\\
&s.t.&r_i = \sum_{j=1}^{M}\lambda_{i,j}R_{i,j} \hspace{0.1in} \forall i \in \N \nonumber\\
&&\sum_{i=1}^{N}\lambda_{i,j} \leq 1 \hspace{0.4in}  \forall j\in \M \nonumber\\
&{\text{variables:}}&\lambda_{i,j} \geq 0 \hspace{0.37in}  \forall i\in\N, j\in\M\nonumber\\
\nonumber
\end{eqnarray}
\vspace{-0.3in}

By capturing the first constraint in the objective function we can reformulate $\mathcal{P}_2$ as

\vspace{-0.1in}
\begin{eqnarray}
&\mathcal{P}_3: \hspace{0.1cm} \max &\sum_{i=1}^{N} (\w_i\log(\sum_{j=1}^{M}\lambda_{i,j}R_{i,j}))  \nonumber\\
&s.t.&\sum_{i=1}^{N}\lambda_{i,j} \leq 1 \hspace{0.4in}  \forall j\in \M \nonumber\\
&{\text{variables:}}&\lambda_{i,j} \geq 0 \hspace{0.37in}  \forall i\in\N, j\in\M\nonumber\\
\nonumber
\end{eqnarray}
\vspace{-0.3in}

We can use dual decomposition to solve $\mathcal{P}_3$ since the constraints that couple the $\lambda_{i,j}$ variables (\ie, the first line of constraints in $\mathcal{P}_3$) can be relaxed using Lagrange duality, and then the optimization problem decouples into several subproblems that as we show next can be solved distributedly. 

Let $\mu_j$ be the Lagrange multiplier for the $j^{th}$ constraint. Then the Lagrangian of $\mathcal{P}_3$ can be written as

\begin{equation}
\begin{split}
L(\blambda,\bmu) = \sum_{i=1}^{N} (\w_i\log(\sum_{j=1}^{M}\lambda_{i,j}R_{i,j})) + \sum_{j=1}^{M}\mu_j(1-\sum_{i=1}^{N}\lambda_{i,j})\\
= \sum_{i=1}^{N} \Bigg[\w_i\log(\sum_{j=1}^{M}\lambda_{i,j}R_{i,j}) - \sum_{j=1}^{M}\mu_j\lambda_{i,j}\Bigg] + \sum_{j=1}^{M}\mu_j
\end{split}
\end{equation}

Here $\blambda$ is the vector of original optimization variables, which are also referred to as primal variables. The Lagrange multipliers ($\mu_j$) are also referred to as dual variables. The problem now separates into two levels of optimization~\cite{MUNGNUM}. At the lower level, each client $i$ needs to solve the following Lagrangian subproblem for a given $\bmu$

\vspace{-0.1in}
\begin{eqnarray}
\label{eq:Lag-subproblem}
& \max\limits_{\lambda_{i,j}}\hspace{0.1in}& \w_i\log(\sum_{j=1}^{M}\lambda_{i,j}R_{i,j}) - \sum_{j=1}^{M}\mu_j\lambda_{i,j}  \nonumber\\
&s.t.& \lambda_{i,j} \geq 0 \hspace{0.37in}  \forall i\in\N, j\in\M \\
\nonumber
\end{eqnarray}
\vspace{-0.3in}

At a higher level, we have the master dual problem in charge of updating the dual variables ($\bmu$) by solving the following dual problem:

\vspace{-0.1in}
\begin{eqnarray}
\label{eq:dual}
& \min\limits_{\bmu}\hspace{0.1in}& \sum_{i}g_i(\bmu) +\sum_{j=1}^M\mu_j\nonumber\\
&s.t.& \bmu \geq \textbf{0} \\
\nonumber
\end{eqnarray}
\vspace{-0.3in}

\noindent
where $g_i(\bmu)$ is the dual function, obtained as the maximum value of the Lagrangian subproblem solved in (\ref{eq:Lag-subproblem}) for a given $\bmu$. This approach solves the dual problem. However, since the original problem in $\mathcal{P}_3$ is convex (and there exists a strictly feasible solution), solving the dual problem equivalently solves the primal problem in $\mathcal{P}_3$. 

Note that the objective function in (\ref{eq:dual}) is convex and differentiable. Hence, we can use the following simple gradient method at each BS $j$ to solve (\ref{eq:dual}):

\begin{equation}
\label{eq:step-adaptation}
\mu_j(t+1) = \Bigg[\mu_j(t) - \gamma \Big[1-\sum_{i=1}^{N}\lambda^{*}_{i,j}(t)\Big]\Bigg]^{+}
\end{equation}

\noindent
where $\lambda^{*}_{i,j}$ is the solution to (\ref{eq:Lag-subproblem}), $t$ is the iteration index, $\gamma > 0$ is a positive step size, and $[.]^{+}$ denotes the projection into the non-negative orthant.

As $t\rightarrow\infty$, the dual variables converge to the dual optimal $\bmu^{*}$ and the primal variables $\blambda^{*}(\bmu(t))$ converge to the optimal primal variable $\blambda^{*}$. Algorithm DDNUM shown below, summarizes the above steps.

\pagebreak
\noindent\rule{9cm}{1pt}\\
\textbf{DDNUM: Dual Decomposition Based Resource Allocation}\\
\noindent\rule{9cm}{1pt}
\textbf{Inputs:} Known $R_{i,j}$ at each client $i$ for every BS $j$ for which $R_{i,j} > 0$.\\
\textbf{Initialization:} Set $t = 0$ and $\bmu(0)$ to some nonnegative value for each BS.
\begin{itemize}
\compactlist


\item {\bf{Step 1:}} Each client $i$ locally solves its Lagrangian problem in (\ref{eq:Lag-subproblem}), $\ie$, finds its time fractions ($\lambda^{*}_{i,j}(\bmu(t))$) for each BS $j$ with $R_{i,j}>0$, and informs those BSs.


\item {\bf{Step 2:}} Each BS updates its price according to Eq.~(\ref{eq:step-adaptation}) and broadcasts the new price to all its clients (\ie, clients with $R_{i,j} > 0$).

\item {\bf{Step 3:}} Set $t \leftarrow t+1$ and go to step 1 (until the satisfying termination point is reached).

\end{itemize}

\noindent\rule{9cm}{1pt}

\end{document}